\documentclass[useAMS,usenatbib]{mnras}
\usepackage[T1]{fontenc} 
\usepackage{aecompl}
\usepackage{amsmath}
\usepackage{amssymb}
\usepackage{graphicx}
\usepackage{booktabs}
\def\refjnl#1{{\rm#1}}
\def\aj{\refjnl{AJ}}                   % Astronomical Journal
\def\araa{\refjnl{ARA\&A}}             % Annual Review of Astron and Astrophys
\def\apj{\refjnl{ApJ}}                 % Astrophysical Journal
\def\apjl{\refjnl{ApJ}}                % Astrophysical Journal, Letters
\def\apjs{\refjnl{ApJS}}               % Astrophysical Journal, Supplement
\def\aap{\refjnl{A\&A}}                % Astronomy and Astrophysics
\def\aaps{\refjnl{A\&AS}}              % Astronomy and Astrophysics, Supplement
\def\mnras{\refjnl{MNRAS}}             % Monthly Notices of the RAS
\def\pasa{\refjnl{PASA}}               % Publications of the Astron. Soc. of Australia
\def\pasp{\refjnl{PASP}}               % Publications of the ASP
\def\rmxaa{\refjnl{Rev. Mexicana Astron. Astrofis.}}%% Revista Mexicana de Astronomia y Astrofisica

\newcommand{\mach}{$\mathcal{M}$} 
\newcommand{\Z}{$Z$} 
\newcommand{\tff}{$t/t_{\rm ff}$} 
\newcommand{\g}{$G_0$}

\newcommand{\CII}{C$\, \scriptstyle\rm II$}

\newcommand{\simgt}{\lower.5ex\hbox{\gtsima}} 
\newcommand{\simlt}{\lower.5ex\hbox{\ltsima}} 
\newcommand{\simpr}{\lower.5ex\hbox{\prosima}}   
\newcommand{\gtsima}{$\; \buildrel > \over \sim \;$} 
\newcommand{\ltsima}{$\; \buildrel < \over \sim \;$} 
\newcommand{\msun}{\,{\rm M_\odot}}
\newcommand{\zsun}{\,{\rm Z_\odot}}
\newcommand\textlcsc[1]{\textsc{\MakeLowercase{#1}}}
\def\cloudy{\textlcsc{cloudy}}
\def\starburst99{\textlcsc{starburst99}}

\title[CO line emission at high-z]{CO line emission from galaxies in the Epoch of Reionization}
\author[Vallini et al.]{L. Vallini$^{1}$\thanks{E-mail: livia.vallini@su.se}, A. Pallottini $^{4,2,5,6}$, A. Ferrara$^{2,3}$, S. Gallerani$^{2}$, E. Sobacchi$^{7,8}$, C. Behrens$^{2}$\\ 
$^1$Nordita, KTH Royal Institute of Technology and Stockholm University, Roslagstullsbacken 23, SE-10691 Stockholm, Sweden\\
$^2$Scuola Normale Superiore, Piazza dei Cavalieri 7, I-56126, Pisa, Italy\\
$^3$Kavli IPMU (WPI), Todai Institutes for Advanced Study, the University of Tokyo, Japan\\
$^4$Centro Fermi, Museo Storico della Fisica e Centro Studi e Ricerche ``Enrico Fermi'', Piazza del Viminale 1, Roma, 00184, Italy\\
$^5$Kavli Institute for Cosmology, University of Cambridge, Madingley Road, Cambridge CB3 0HA, UK\\
$^6$Cavendish Laboratory, University of Cambridge, 19 J. J. Thomson Ave., Cambridge CB3 0HE, UK\\
$^7$Physics Department, Ben-Gurion University, P.O.B. 653, Beer-Sheva 84105, Israel\\
$^8$Department of Natural Sciences, The Open University of Israel, 1 University Road, P.O.B. 808, Raanana 4353701, Israel
}
\date{}
\pubyear{2017}

\begin{document}
\label{firstpage}
\pagerange{\pageref{firstpage}--\pageref{lastpage}} 
\maketitle
\begin{abstract}
We study the CO line luminosity ($L_{\rm CO}$), the shape of the CO Spectral Line Energy Distribution (SLED), and the value of the CO-to-$\rm H_2$ conversion factor in galaxies in the Epoch of Reionization (EoR). To this aim, we construct a model that simultaneously takes into account the radiative transfer and the clumpy structure of giant molecular clouds (GMCs) where the CO lines are excited. We then use it to post-process state-of-the-art zoomed, high resolution ($30\, \rm{pc}$), cosmological simulation of a main-sequence ($M_{*}\approx10^{10}\, \rm{M_{\odot}}$, $SFR\approx 100\,\rm{M_{\odot}\, yr^{-1}}$) galaxy, ``Alth{\ae}a'', at $z\approx6$. We find that the CO emission traces the inner molecular disk ($r\approx 0.5 \,\rm{kpc}$) of Alth{\ae}a with the peak of the CO surface brightness co-located with that of the [\CII] 158$\rm \mu m$ emission. Its $L_{\rm CO(1-0)}=10^{4.85}\, \rm{L_{\odot}}$ is comparable to that observed in local galaxies with similar stellar mass.
The high ($\Sigma_{gas} \approx 220\, \rm M_{\odot}\, pc^{-2}$) gas surface density in Alth{\ae}a, its large Mach number (\mach$\approx 30$), and the warm kinetic temperature ($T_{k}\approx 45 \, \rm K$) of GMCs yield a CO SLED peaked at the CO(7--6) transition, i.e. at relatively high-$J$, and a CO-to-$\rm H_2$ conversion factor $\alpha_{\rm CO}\approx 1.5 \, \rm M_{\odot} \rm (K\, km\, s^{-1}\, pc^2)^{-1} $ lower than that of the Milky Way. The ALMA observing time required to detect (resolve) at 5$\sigma$ the CO(7--6) line from galaxies similar to Alth{\ae}a
is $\approx13$ h ($\approx 38$ h).
\end{abstract}

%%%% INTRODUCTION %%%%%%%%%
\section{Introduction}\label{introduction}
Constraining the properties of the molecular gas in galaxies at the end ($z\approx 6$) of the Epoch of Reionization (EoR) is a compelling step to understand the process of star formation in the first galaxies.

Molecular hydrogen ($\rm H_2$), the most abundant molecule in the Universe, lacks of a permanent dipole moment and its first quadrupole line has an excitation temperature ($T_{ex}\approx500$ K) significantly higher than the kinetic temperatures ($T_{K}\approx10-20$ K) of giant molecular clouds (GMCs) \citep{mckee2007}. This is the reason why molecular gas in galaxies is very often traced through the detection of the rotational transitions the carbon monoxide, CO, the second most abundant molecule after $\rm H_2$. The first CO rotational transition is in fact characterised by $T_{ex}\approx 5$ K, with critical density $n_{cr}\approx 10^{3}\,\rm{cm^{-3}}$ \citep[][]{solomon2005, carilli2013}, i.e. it is easily excited within GMCs.

Noticeably, the different excitation requirements of the various CO lines ($T_{ex}\approx 5-300$ K, $n_{cr}\approx 10^{3}-10^{6}\,\rm{cm^{-3}}$ for upper state rotational quantum number $J_{\rm up}=1-10$) can be exploited to constrain gas properties (e.g. density, temperature), and the gas heating mechanisms (e.g. FUV photons, X-ray photons, cosmic rays, shocks). This can be done through the analysis of the so-called CO Spectral Line Energy Distribution (CO SLED) - flux in each emission line as a function of $J_{\rm up}$ \citep[][]{kaufman1999, meijerink2007, obreschkow2009, mashian2015, rosenberg2015, lu2017, indriolo2017}.

Searches for CO line emission at redshift $z>5$  have been mainly focused on the most luminous sources such as QSOs \citep[e.g.][for a recent review]{bertoldi2003, maiolino2007, wang2010, walter2012, combes2012, venemans2012, gallerani2014, gallerani2017} or powerful sub-millimeter galaxies \citep[e.g.][]{riechers2010, weiss2013, aravena2016}. On the contrary, little is known regarding the molecular gas content of high-$z$ normal star-forming galaxies -- e.g. Lyman Alpha Emitters (LAEs) and/or Lyman Break Galaxies (LBGs) -- which are more representative of the bulk of galaxy population at the end of EoR \citep[e.g.][]{dayal2008,dayal2009, vallini2012}. Only a handful of CO detections have been reported in (mostly lensed) LBGs at $z\approx3$ \citep{baker2004, coppin2007, riechers2010b,  livermore2012, saintonge2013, dessauges-zavadsky2015, dessauges-zavadsky2016, ginolfi2017}, while searches for low-$J$ CO rotational lines in LAEs at $z\approx6$ have resulted in non-detections \citep{wagg2009, wagg2012}.

The ALMA advent has opened new perspectives for the detection of CO rotational transitions from the EoR. It is more sensitive than the other submillimeter/millimeter facilities adopted in previous studies targeting LAEs/LGBs in the EoR and, more importantly, its receiver bands allow one to trace transitions with $J_{\rm up}\geq 6$ (high-$J$, hereafter), from $z\geq 6$. This represents a key-point because the peak of the CO SLED in galaxies with high specific star formation rate (sSFR) is often associated with high-$J$ transitions \citep{mashian2015}, and such galaxies are common at high-$z$ \citep{jiang2016}. Those transitions, being more luminous than low-$J$ ones, may represent a viable option to target molecular gas at the end of EoR.

The advent of ALMA has also triggered the development of -- mostly semi analytical -- works that aim at modelling the CO emission signal from high-redshift ($z>6$) galaxies \citep[e.g.][]{obreschkow2009,lagos2012, vallini2012, munoz2013, dacunha2013, popping2016}. 

\citet{obreschkow2009, lagos2012} and \citet{popping2016} combined semi analytical galaxy formation simulations with sub-grid models devised to convert the molecular mass into CO luminosities, and predict the high-$z$ evolution of the CO luminosity function \citep[e.g.][]{walter2014, decarli2016, vallini2016}. Those models account for e.g. the heating by the cosmic microwave background (CMB) \citep[see also][]{dacunha2013}, the Far-Ultraviolet (FUV) flux from starburst, and/or by X-rays produced by active galactic nuclei. \citet{vallini2012} used the semi-analytical model for the molecular fraction by \citet{krumholz2009} to infer the CO luminosity from a sample of $z\approx6$ LAEs extracted from cosmological simulations. However, the work by \citet{obreschkow2009, lagos2012, vallini2012} lacked of a detailed description of the internal structure of GMCs. \citet{munoz2013} made a step forward in this direction. Using an analytical model, they explored the link between the GMC properties set by the turbulence, and the physics of CO emission in high-$z$ LBGs; they conclude that the CO signal could be very difficult to observe from $z>6$. While \citet{munoz2013} catch some of the fundamental aspects of the internal structure and of the radiative transfer within the GMCs, this was at the expense of a full description of the galaxy formation process.

The goal of this work is to assess the feasibility of detecting CO lines from typical LBGs at high-$z$ by simultaneously capturing the full cosmological context of high-redshift galaxy formation, and the radiative transfer from the outer photodissociation regions (PDRs) \citep{hollenbach1999} up to the fully molecular inner part of GMCs. To this aim, we construct a physically motivated model that simultaneously takes into account the radiative transfer and the clumpy structure of GMCs. We then use it to post-process state-of-the-art zoomed cosmological simulations \citep[][]{pallottini2017, pallottini2017b}.
This type of sub-grid approach has been already shown to be an optimal strategy to obtain predictions and insights on the luminosity of, e.g., [\CII] 157.7$\rm \mu m$ line emission tracing the neutral diffuse gas and dense PDRs \citep[][]{vallini2013,vallini2015, pallottini2015} in galaxies at the end or EoR.
It allows one, on the one hand, to properly treat the small physical scales ($\approx 0.1-1$ pc) of clumps in GMCs, and the complex network of chemical and physical reactions in the PDR layer and in the fully molecular parts of GMCs. On the other hand, it benefits from the high-resolution hydrodynamical simulations by obtaining a proper description, down to scales of $\approx 30$ pc, of the ISM density, turbulence level, metal enrichment, and radiation field into which GMCs are embedded.
 
The paper is structured as follows: in Sec. \ref{outline} we describe how we implement the CO line emission calculation taking into account the internal density structure of GMCs.
In Sec. \ref{validation} we validate the model against local observations. Finally, in Sec. \ref{co_highz}, we apply the model to high-resolution cosmological simulations to compute the CO emission from normal star-forming galaxies at the end of EoR.
We draw our conclusions in Sec. \ref{conclusions}.

\section{Model outline}\label{outline}
The model strategy adopted in this work is summarised in Fig. \ref{model_outline}, where we outline its modular structure. The first part (see Sec. \ref{GMC_modelling}) deals with the analytical description of the \textit{internal density structure} of GMCs and its time evolution. The second part concerns the \textit{radiative transfer} performed to compute the CO line emission once the density field is established (see Sec. \ref{rt_modelling}). 
The sub-grid model is optimised for implementation in cosmological simulations able to approximately resolve GMC scales \citep[$\approx 30-100$ pc, see][]{mckee2007}.

\subsection{The internal structure of GMCs}\label{GMC_modelling}
Fragmentation of gas clouds has been intensively studied in the past, and both numerical and analytical models \citep[e.g][]{vazquez-semadeni1994, krumholz2005, padoan2011, hennebelle2011, hennebelle2013, kim2003, wada2008, tasker2009, federrath2013} have shown that the density field of an isothermal, non-gravitating, turbulent gas of mean density $\rho_0$ is well described by a log-normal probability distribution function (PDF). The volume-weighted PDF ($P_V(\rho)$) can be written as:
\begin{equation}\label{lognormalfunction}
P_V(\rho) = \frac{1}{(2 \pi \sigma^2)^{1/2}}\, {\rm exp} \left[ -\frac{1}{2} \left( \frac{{\rm ln}(\rho/\rho_0)- \langle \rm ln(\rho/\rho_0) \rangle}{\sigma} \right)^2 \right],
\end{equation}
where $\rho_0$ is the mean cloud density, and the volume-averaged value of the logarithm of the density is related to $\sigma$ by $\langle \rm ln(\rho/\rho_0) \rangle=- \sigma^2/2$ \citep[e.g][]{ostriker2001, federrath2013}. The latter depends on the sonic Mach number ($\mathcal{M}$) through the following relation:
 \begin{equation}\label{lognormaleq}
\sigma^2={\ln}\left( 1+ b^2 \mathcal{M}^2 \right),
\end{equation}
where $b$ parametrises the kinetic energy injection mechanism (often referred to as forcing) driving the turbulence \citep[$b\approx0.3-1$, see][]{molina2012}. Throughout this paper we assume $b=0.3$.
\begin{figure}
\centering
\includegraphics[scale=0.32]{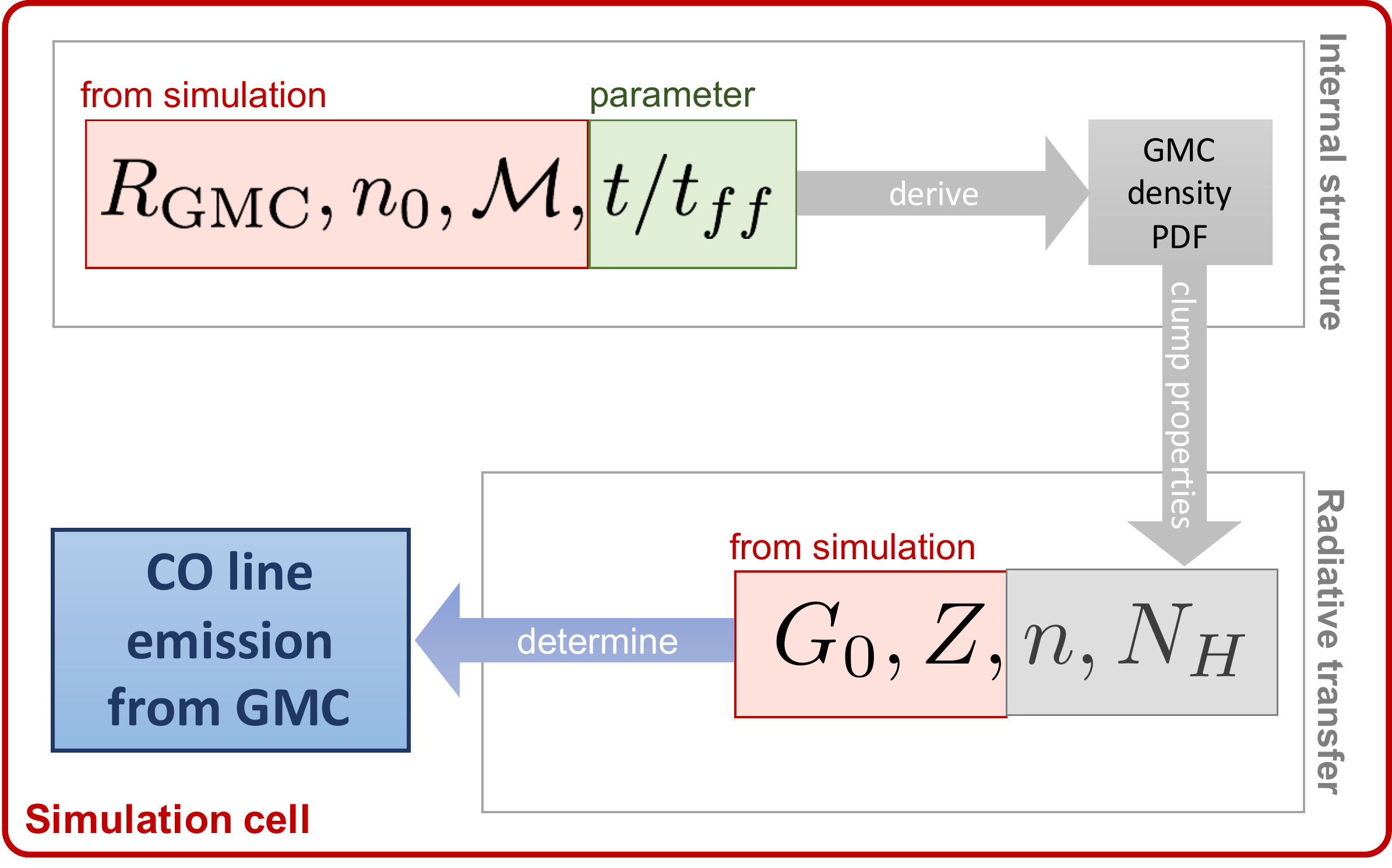}
\caption{Outline of the model presented in this paper. The modelling of the GMC structure as a function of the cloud radius ($R_{\rm GMC}$), mean number density ($n_0$), sonic Mach number (\mach), and time (\tff) is described in details in Sec. \ref{GMC_modelling}. The radiative transfer, as a function of FUV flux in Habing units ($G_0$), metallicity ($Z$), gas density ($n$), and column density ($N_H$) is discussed Sec. \ref{rt_modelling}. The application of the sub-grid model to simulations is presented in Sec \ref{coupling_subgrid}. \label{model_outline}}
\end{figure}

When self-gravity becomes important, the probability of finding dense regions increases, and a power-law tail develops on the high-density side of the PDF. The occurrence of the power-law tail is confirmed both theoretically \citep[e.g.][]{krumholz2005, hennebelle2011, padoan2011, federrath2013}, and observationally via molecular line detections \citep[e.g.][]{goldsmith2008, goodman2009, schneider2016} or dust extinction measurements carried out in nearby GMCs \citep[e.g.][]{kainulainen2009, lombardi2015,stutz2015, schneider2016}.
Treating properly the high-density tail of the density PDF is pivotal when computing the emission of high-$J$ CO rotational lines as they have high critical densities ($n_{crit}\geq2.9 \times 10^5\rm cm^{-3}$ for $J_{up}\geq6$), and trace the densest regions of GMCs \citep[e.g.][and references therein]{carilli2013}. 

In our model we describe the time evolution of the density PDF of self-gravitating GMCs via the formalism developed by \citet{girichidis2014}. Assuming a pressure-free collapse, these authors provide a set of analytical equations to calculate the functional form of the PDF at any given time, $P_V(\rho,t)$, given the initial $P_V(\rho,0)$. According to their model, the high-density tail of the PDF quickly asymptotes to a power-law, consistently with observations. Details on the calculation of $P_V(\rho,t)$ from the initial $P_V(\rho,0)$ are given in Appendix \ref{sec:ghirichidismodel}.

As the initial PDF, in this paper we take $P_V(\rho,0)\equiv P_V(\rho)$, where $P_V(\rho)$ is defined in eq. \ref{lognormalfunction}. The density at which $P_V(\rho,t)$ starts to deviate from the lognormal, $\rho_{\rm tail}$, moves with time to lower densities. This is due to the fact that collapse at a given density $\rho$ can only manifest itself after approximately a free-fall time, $t_{\rm ff}(\rho)$ (see eq. \ref{app_tff}), implying that the collapse of lower density parcels takes a longer time. \citet{girichidis2014} find that the tail is well defined above a density $\rho_{\rm tail}(t)$ that evolves with time according to the following relation:
\begin{equation}
\frac{\rho_{\rm tail}(t)}{\rho_0} \approx 0.2 \left( \frac{t}{t_{\rm ff}(\rho_0)} \right)^{-2.0}.
\end{equation}
This the reason why (see the schematic in Fig. \ref{model_outline}), the density PDF of a GMC is ultimately function of two parameters: (i) the mach number (\mach) -- which affects the lognormal part of the distribution --, and (ii) the ratio (\tff) -- that determines the point at which the PDF significantly deviates from the lognormal distribution.
Note that the additional time-dependence of the density PDF due to FUV-photoevaporation \citep{gorti2002, vallini2017, decataldo2017} is neglected in this work.

Let now consider a GMC of radius $R_{\rm GMC}$, volume $V_{\rm tot}=4/3 \pi R_{\rm GMC}^3$, and mean density\footnote{Unless otherwise stated we consider a mixture of neutral hydrogen and helium characterised by a mean molecular weight $\mu=1.4$} ($\rho_0=n_0 \mu m_p$), characterised by a fixed Mach number \mach~ and an evolutionary time \tff. The normalisation of the volume-weighted density PDF must satisfy the following relation:
\begin{equation}
V_{\rm tot}= \int P_V(\rho | \mathcal{M}, t/t_{\rm ff}) \rm{d}\rho,
\end{equation}
which allows us to associate to each density $\rho_i$ a typical length scale,
\begin{equation}
\label{clump_radius}
r_{i}=\left(V_{\rm tot} \int_{\rho_i - \delta_i}^{\rho_i + \delta_i} P_V(\rho | \mathcal{M}, t/t_{\rm ff}) \rm{d}\rho \right)^{1/3},
\end{equation}
and column density $N_i (\rho_i)= ({\rho_i}/{\mu m_p}) r_i$. The CO emission per unit volume from each density element $\rho_i$ can be computed as follows:
\begin{equation}
\label{columinosity}
l_{\rm CO, J}(\rho_i) =\frac{1}{V_{tot}} {\varepsilon}_{\rm CO, J}(n_i, N_i, Z, G_0)\, 4 \pi r_i^2,
\end{equation}
where $ {\varepsilon}_{\rm CO, J}(n_i, N_i, Z, G_0)$ is the CO emissivity of the $J\rightarrow J -1 $ transition as a function of the gas element (i) density, $n_i$,  (ii) column density, $N_i$, (iii) metallicity, $Z$, and (iv) Far-Ultraviolet (FUV) flux, $G_0$, in the Habing band ($6-13.6 \, \rm{eV}$) normalised to that in the solar neighbourhood \citep[$\approx 1.6\times 10^{-3}\rm {erg\, cm^{-2}\, s^{-1}}$][]{habing1968}. The total CO emission\footnote{Our treatment implicitly assumes that the each clump is exposed to the same imposed external FUV flux. This assumption neglects possible radiation anisotropies depending on the position of the clump within the GMC, and shadowing effects \citep[see Appendix A][]{vallini2017}.} from the GMC is then:
\begin{equation}
\label{cotot}
L^{tot}_{\rm CO,J} = \int l_{\rm CO, J} (\rho) P_V(\rho | \mathcal{M}, t/t_{\rm ff}) \rm{d} \rho. 
\end{equation}
It is useful to express eq. \ref{cotot} in terms of $s={\rm ln}(\rho/\rho_0)$:
\begin{equation}\label{lco_of_s}
 L^{tot}_{\rm CO,J} = \int \mathcal{L}_{\rm CO, J}(s) \mathrm{d}s,
\end{equation}
where $\mathcal{L}_{\rm CO, J}=l_{\rm CO, J} P_V/\rho$.
\subsection{Radiative transfer}\label{rt_modelling}
We calculate the value of $\varepsilon_{\rm CO, J}$ for the first 9 rotational transitions of the CO molecule with the \cloudy~ version c13.03 \citep{ferland2013}. 

\cloudy~includes $\approx1000$ reactions involving molecules containing H, He, C, N, O, Si, S, and Cl atoms \citep[see Appendix A of][for details on the molecular network]{abel2005}. Details on the CO network, including CH, CH$^+$, OH, OH$^+$, $\rm H_2O$, $\rm H_2O^+$, $\rm H_3O^+$, $\rm O_2$ and $\rm O_2^+$ are presented in \citet{ferland1994}. The majority of reaction rates come from the UMIST 2000 database \citep{leteuff2000}. The treatment of the formation and dissociation of the $\rm H_2$ molecule is outlined in \citet{shaw2005} and it accounts for $\rm H_2$ formation via gas-phase reactions, and on the dust grain surface \citep{cazaux2004}. The local grain properties (temperature, charge) at each point in the cloud are computed self-consistently. \cloudy~also treats the primary and secondary cosmic-ray (CR) ionisation processes. We adopt the default CLOUDY prescriptions for the CR ionization rate background ($\zeta_{\rm CR}$; CRIR, hereafter) $\zeta_{\rm CR} = 2\times10^{-16} \rm s^{-1}$ \citep{indriolo2007}. The $\rm H_2$ secondary ionisation rate is $=4.6\times10^{-16}\rm s^{-1}$ \citep{glassgold1974}. As the CRIR is a fundamental parameter that may have strong effects on the gas temperature and chemical composition at high densities, we discuss the impact of our assumption on the CO luminosity inferred with our model in the Appendix \ref{appendix_CRs}.

We adopt a 1D geometry, assuming a gas slab of density $n$, and fixed metallicity $Z$. The spectral energy distribution (SED) of the radiation field impinging on the slab surface is calculated using the stellar population synthesis code \starburst99 \citep{leitherer1999}, assuming a Salpeter Initial Mass Function in the range $1-100\,\rm M_\odot$, and considering a continuous star formation mode with an age of the stellar population of 10 Myr. We adopt the Geneva standard evolutionary tracks \citep{schaller1992} with metallicity $Z_*=\rm{1\, \zsun} ,\, 0.2 \,\zsun, 0.05 \,\rm{\zsun}$, and Lejeune-Schmutz stellar atmospheres which incorporate plane-parallel atmospheres and stars with strong winds \citep{lejeune1997, schmutz1992}. The SED is normalised so that the Habing flux varies in the range $G_0=10^0-10^{4.5}$.

We adopt the gas-phase abundances ($\rm C/H=2.51 \times 10^{-4}$, $\rm O/H=3.19 \times 10^{-4}$, $\rm N/H=7.94 \times 10^{-5}$, $\rm S/H =3.24 \times 10^{-5}$)\footnote{For the complete list of the abundances please refer to \citet{ferland2013} and to the \cloudy~ manual Hazy I.} provided in \cloudy \, for the ISM of the Milky Way. The abundances are an average of those measured in the cold and warm diffuse phases of the galactic ISM by \citet{cowie1986} and \citet{savage1996}. 

We implement the \citet{weingartner2001} grain size distribution ($\rm{d}n_{gr}/\rm{d}a$; GSD) which has been shown to reproduce (among others) the SMC extinction curve. This study provides a set of analytical equations (see eqs. 4--6, and the best fit parameters in their Tab. 3) that (i) imposes a smooth cutoff for sizes greater than a threshold $a_t$, (ii) controls the steepness of this cutoff, and (iii) allows for a change in the slope ($\rm{d ln}\, n_{gr}/\rm{d ln}\, a$) below $a_t$.

As we are interested in obtaining predictions on the CO luminosity in high-$z$ galaxies, we set the floor temperature of our \textit{fiducial} simulations to that of CMB at $z=6$, i.e. $T_{\rm CMB}=2.73 (1+z) {\rm K}=19.1$ K. The line emission from molecular gas is affected in two ways by the CMB: (i) the higher $T_{\rm CMB}$ leads to an increase of the line excitation, and thus of the line luminosities; (ii) the background against which the line is measured also increases \citep[e.g][]{obreschkow2009, dacunha2013}. To test the behaviour of our model on local observations, and to study the impact of the CMB temperature on our predictions, we run also a set of cases using the present-day CMB temperature, $T_{\rm CMB}=2.73\,\rm{K}$ (see Sec. \ref{study_cmb_and_tail} for a discussion on this point).

We run a total of $500\,(10\times 10 \times 5)$ \cloudy \, simulations varying (in 0.5 dex steps) ${\rm log}(n/\rm{cm^{-3}})$ in the range $[1.5, 6]$, ${ \rm log} \,G_0$ in $[0, 4.5]$, and ${\rm log}(Z/\zsun)$ in $[-2, 0]$, scaling the gas-phase abundances and the dust-to-gas ratio with the metallicity of each specific model.

Such parameter space brackets the plausible range of GMC parameters relevant to high-$z$ in galaxies. 
The code computes the radiative transfer through the slab up to a hydrogen column density $N_{\rm H}=10^{23}\,\rm{cm^{-2}}$. This stopping criterion is chosen to fully sample the molecular part of the illuminated slab, typically located at $N_{\rm H}\simgt 2\times10^{22}\,\rm{cm^{-2}}$ \citep{mckee2007}. 
The output of each run is the CO line emissivity of each $J \rightarrow J - 1$ transition:
\begin{equation}\label{COemissivity}
\varepsilon_{\rm CO, J} = \varepsilon_{\rm CO, J}(n_i, N_i, Z, G_0)
\end{equation}
which enters in eq. \ref{columinosity} and hence in eq. \ref{cotot}. 

\section{Model validation}\label{validation}
\begin{figure} 
\centering
\includegraphics[scale=0.45]{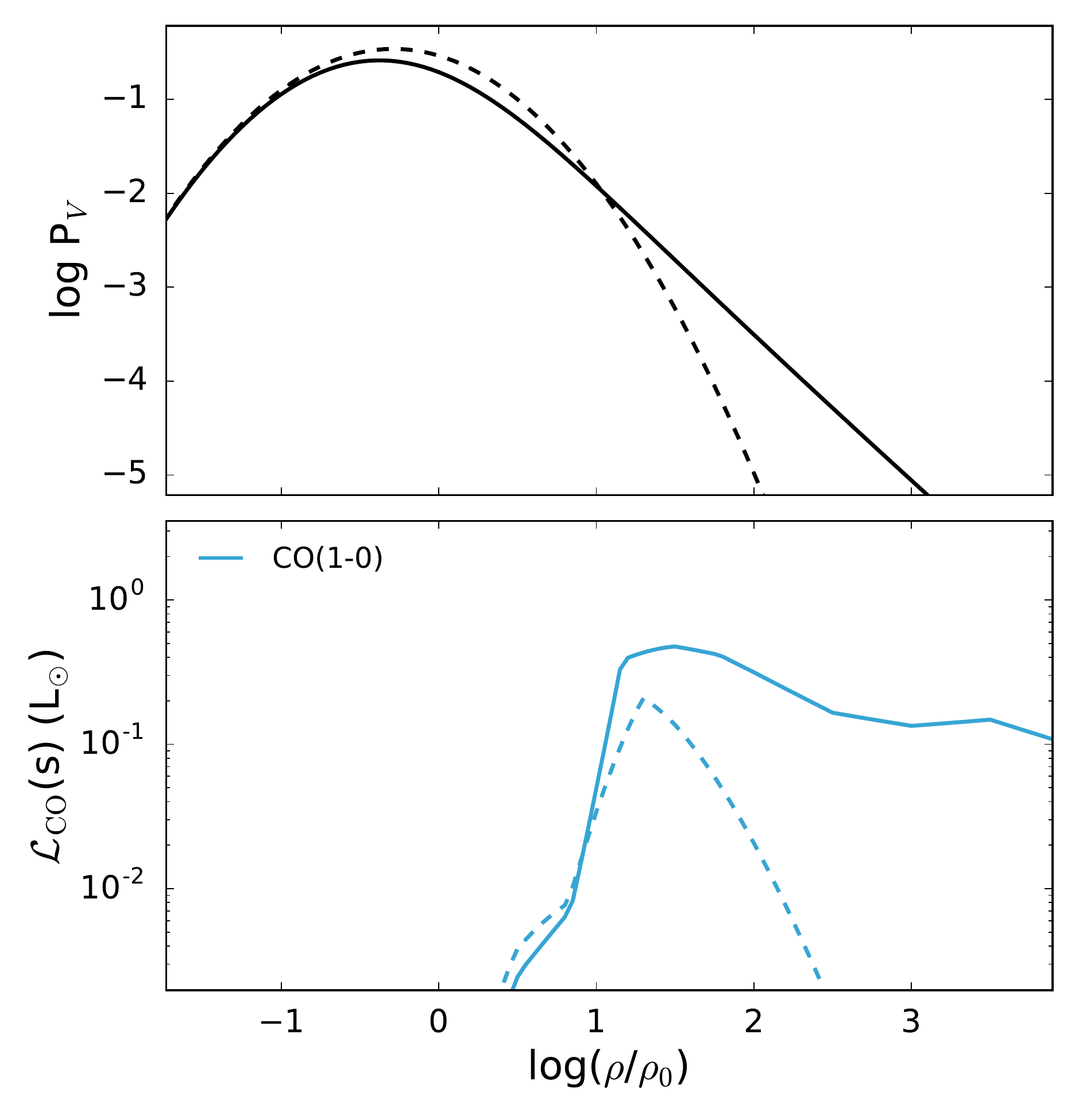}
\caption{Upper panel: initial lognormal volume-weighted density PDF (dashed line,  \mach$= 5$), and the evolved density PDF (lognormal+tail hereafter, solid) after $t/t_{\rm ff}(\rho_0)=0.4$. Lower panel: Specific CO(1--0) luminosity for the initial (dashed) and evolved (solid) cases shown in the upper panel. \label{pdf_Lco}}
\end{figure}

By adopting the procedure described in the previous sections we compute the luminosity of the first 9 CO rotational transitions. The link between the time evolution of the density PDF, and the resulting CO luminosity is illustrated in Figure \ref{pdf_Lco}. In the upper panel, we plot volume-weighted density PDF at $t=0$ (lognormal, \mach$=5$) and the resulting $P_V$ after $t/t_{\rm ff}(\rho_0)=0.4$ (solid line). We choose $t/t_{\rm ff}(\rho_0)=0.4$ to maximise the mass \citep[$97\%$ of the total mass of the GMC,][]{girichidis2014} in the tail, hence highlighting more clearly the effect of the tail appearance on the CO emission.
The \textit{fiducial cloud} (see Tab. \ref{fiducialgmc}) is characterised by: $n_0=100\, \rm{cm^{-3}}$ ($\rho_0=2.3 \times 10^{-22}\, \rm{g\, cm^{-3}}$), $R_{\rm GMC}=15\, \rm pc$, $\rm log(Z/Z_{\odot})=0$, and illuminated by a FUV field with $\rm log\, G_0=2$. In the lower panel we plot the corresponding $\mathcal{L}_{\rm CO}(s)$ (see eq. \ref{lco_of_s}) at $t=0$ (solid lines) and at \tff$=0.4$ (dashed lines). The CO(1--0) emission is boosted at high densities, once the tail has developed at \tff$=0.4$. 

\begin{table*}
\centering
\caption{Parameters of the fiducial GMC.}
\label{fiducialgmc}
\begin{tabular}{@{}cccccc@{}}
\toprule
 $R_{\rm GMC}/(\rm pc)$ & $n_0/(\rm cm^{-3})$ & $\rho_0/\rm (g\, cm^{-3})$ & $M_{\rm GMC}/(\rm M_{\odot}$) & $\rm log(Z/Z_{\odot})$ & $\rm log\, G_0$  \\ \midrule
  $15$            & $100$          &   $2.34\times10^{-22}$  & $4.9\times10^4$ &     $0$        &      $2$     \\
\bottomrule
\end{tabular}
\end{table*}
\subsection{The $M_{\rm vir} -L'_{\rm CO}$ relation}
\label{mvir_lco_section}
Studies of resolved molecular clouds find that GMCs are in approximate virial equilibrium, \citep[e.g.][]{larson1981, solomon1987, bolatto2008} 
and, because of that, they obey scaling relations, ofter referred to as \textit{Larson laws}. Those relations ultimately link the size ($R$), the velocity dispersion ($\sigma$), and the CO luminosity ($L'_{\rm CO}$)\footnote{$L'_{\rm CO}$ (in $\rm K\, km\, s^{-1}\, pc^2$) is linked to $L_{\rm CO}$, (in $L_{\odot}$) through the following relation:  $L_{\rm CO}=3 \times 10^{-11} \nu_r^3 L'_{\rm CO}$, where $\nu_r$ is the rest frequency of the line expressed in GHz \citep{carilli2013}.}of GMCs \citep{solomon1987}:
\begin{eqnarray}
\label{larsonsolomon}
&&\sigma\approx 0.7R^{0.5}\, \rm km\, s^{-1} \nonumber \\
&&L'_{\rm CO}\approx 130 \sigma^5 ~\rm K\, km\, s^{-1} \, pc^2 \\
&&L'_{\rm CO}\approx 25 R^{2.5} ~\rm K \,km \,s^{-1} \,pc^2. \nonumber 
\end{eqnarray}
These relations are valid under the assumption that molecular gas is dominating the mass enclosed in the cloud radius and therefore the virial mass ($M_{\rm vir}$) is a good measure of the $\rm H_2$ traced by CO.\footnote{$M_{vir}=M_{\rm H_2}$ implies $f_{\rm H2}\approx 1$. In Appendix \ref{appendix_molecular_fraction} we compute $f_{\rm H2}$ as a function of  \mach, \g, \Z, and $n_0$. We show that $f_{\rm H2}\approx1$ for a major fraction of the parameter space covered in this study.}

The virial equilibrium implies that $\sigma^2\approx GM_{\rm vir}/R$, thus eqs. \ref{larsonsolomon} translate into the following:
\begin{equation}
\label{m-lco}
M_{\rm vir} \approx 39 L{'}_{\rm CO}^{0.81}\, \rm M_{\odot},
\end{equation}
where we note that the slope of the mass-luminosity relation, $\gamma=0.81$, is the one found by \citet{solomon1987}. Subsequent studies \citep[e.g.][and references therein]{bolatto2008, bolatto2013} have confirmed that $\gamma \approx 1$. 

\begin{figure}
\centering
\includegraphics[scale=0.4]{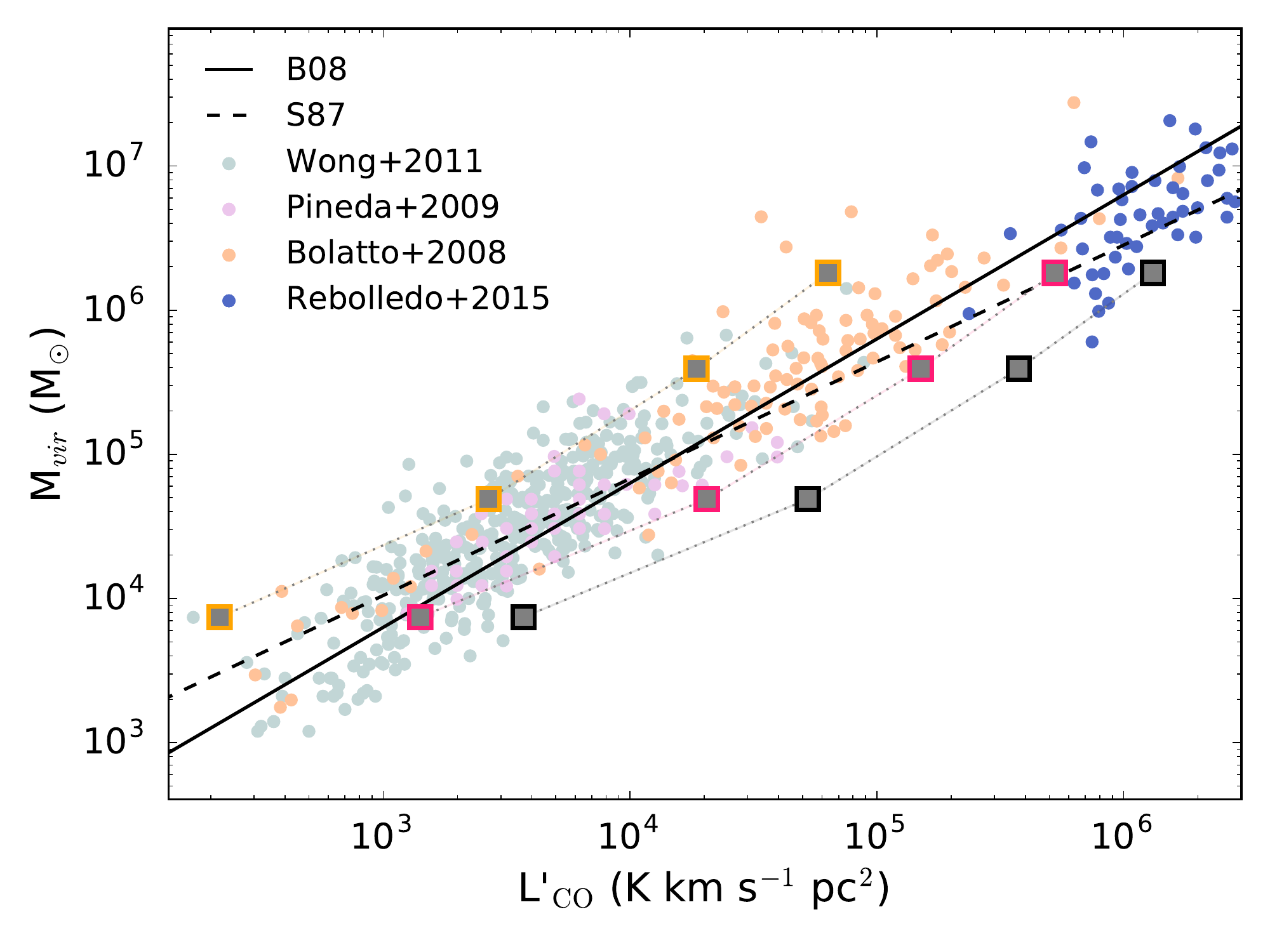}
\caption{CO(1-0) luminosity vs. virial mass of the GMC. The black dashed line is the $M_{\rm vir}$-$L'_{\rm CO}$ relation derived by \citet{solomon1987}, while the black solid line represents the relation found by \citet{bolatto2008}.
The observational data concerning GMCs in nearby galaxies are plotted with coloured points. The orange points represent the compilation by \citet{bolatto2008} including M31, M33, and nine dwarf galaxies. High-resolution observations of GMCs in the LMC by \citet{pineda2009} are plotted with pink points, while LMC data by \citet{wong2011} are indicated with green points. Recent GMCs/molecular complex observations in NGC6946, NGC628, and M101 by \citet{rebolledo2015} are plotted with blue points. 
The $M_{\rm vir}$-$L'_{\rm CO}$ relation resulting from our model is plotted with orange ($Z=0.1\, \zsun$), magenta ($Z=0.5,\, \zsun$), and black ($Z=1\, \zsun$) squares, respectively. \label{mvirlco}}
\end{figure}

In Figure \ref{mvirlco} we plot $M_{\rm vir}-L'_{\rm CO}$ as resulting from our model. More precisely, we calculate the CO(1--0) luminosity fixing the mass ($M_{\rm GMC}$), radius ($R_{\rm GMC}$), considering the $P_V$ at \tff$=0.1$, i.e. when $\approx 50\%$ of the GMC mass is in the PDF tail, and the rest in the lognormal distribution. The Mach number is selected so that it satisfies the following condition:
\begin{equation}
\mathcal{M} c_s= \sqrt \frac{G M_{\rm GMC}}{R_{\rm GMC}}.
\end{equation}
We set the sound speed of the GMC to $c_s(T=10\,\rm{K})\approx 0.3 \, \rm{km\, s^{-1}}$. For the RT we adopt the \cloudy~ runs at $z=0$ (i.e. those with $T_{\rm CMB}=2.73\,\rm{K}$) for different metallicities $Z=1,0.5,0.1\,Z_{\odot}$.
Our results are in nice agreement with observations, and a linear fit between $\rm{log}(L'_{\rm CO})$ and $\rm{log}(M_{\rm vir})$ returns a slope $\gamma=0.96,\,0.92,\,0.93$ for $Z=1,0.5,0.1\,Z_{\odot}$, respectively. The CO luminosity decreases with decreasing $Z$ at fixed $M_{\rm vir}$, implying that the CO-to-$\rm H_2$ conversion factor (see Sec. \ref{alphaco_section}) increases for lower $Z$, as noticed also by e.g. \citet[][]{wolfire2010, glover2011, narayanan2012, bolatto2013}. The model results at $Z=0.5,\, 0.1\, Z_\odot$ enclose the observations by \citet{pineda2009, wong2011} of GMCs in the LMC \citep[$Z_{\rm LMC}\approx 0.5\, Z_{\odot}$][]{rolleston2002, chevance2016, lee2016}.

\subsection{The impact of the high-density tail}\label{study_cmb_and_tail}
\begin{figure}
\centering
\includegraphics[scale=0.4]{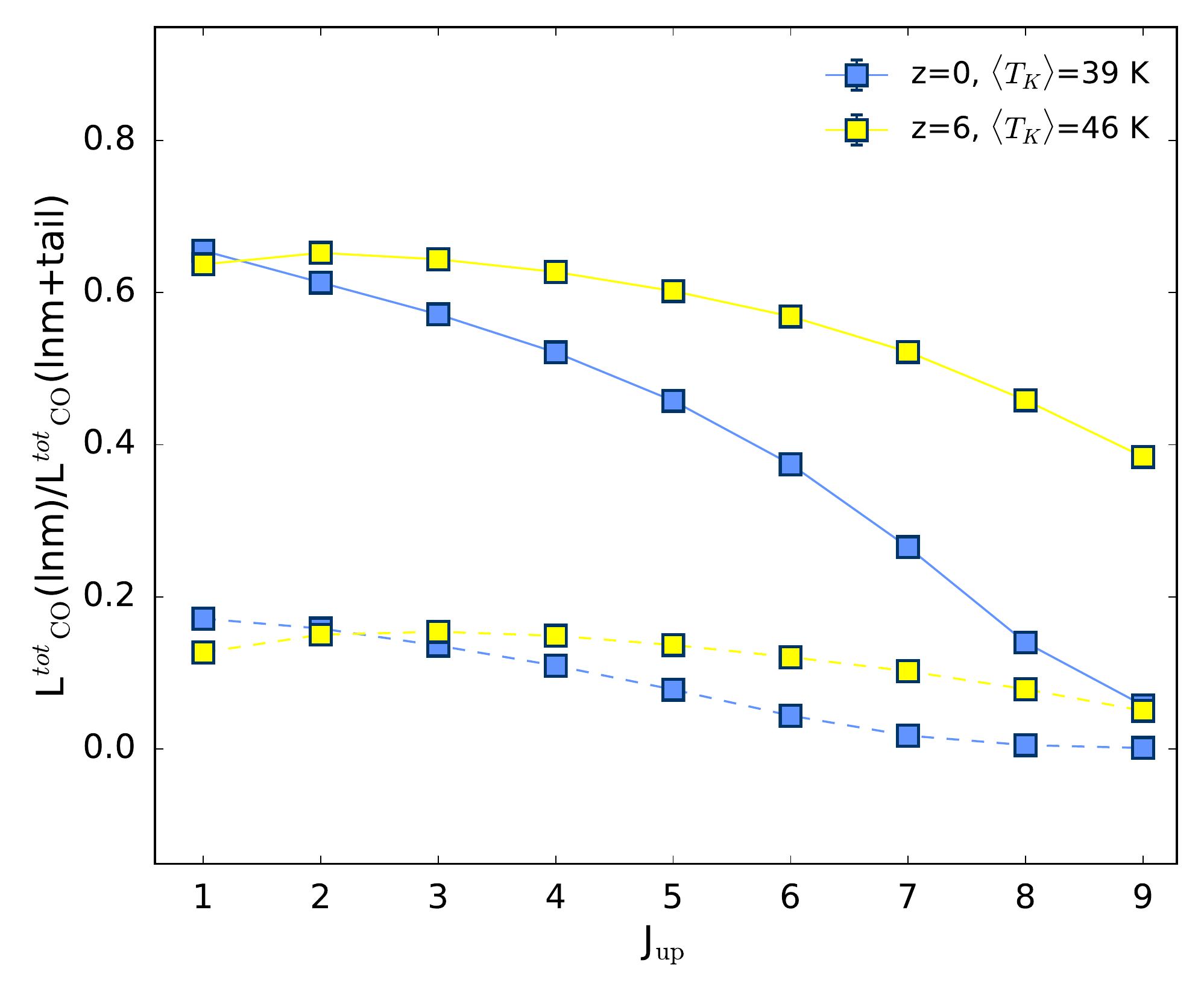}
\caption{Ratio of the total luminosity ($L^{tot}_{\rm CO}$) of the first 9 CO rotational transitions for the fiducial GMC (see Table \ref{fiducialgmc}) at $t=0$ (i.e. for a lognormal $P_V$) to that including the density PDF power-law tail at $t/t_{\rm ff}(\rho_0)=0.1$.  Solid (dashed) lines represent the results obtained for a GMC characterised by \mach$=10$ (\mach$=5$), located at different redshift, $z=0,\, 6$, (blue/yellow respectively). The mean kinetic temperature at $N_H>10^{21.5}\, \rm{cm^{-2}}$ of the GMCat $z=0$ is $\langle T_K \rangle =39\, \rm{K}$, while at $z=6$ is $\langle T_K \rangle =46\, \rm{K}$. \label{ratio_tail}}
\end{figure}
In this Section we discuss how the high-density power-law tail affects the luminosity of the various CO transitions. To do that we compare the pure log-normal density field for the fiducial GMC ($L^{tot}_{\rm CO}$($\rm lnm$)), and a distribution which includes the power-law tail after $t/t_{\rm ff}(\rho_0)=0.1$ ($L^{tot}_{\rm CO}$($\rm lnm+tail$)).

In Figure \ref{ratio_tail} we plot the ratio $R_{\rm CO,J} = L^{tot}_{\rm CO}(\rm lnm)/ L^{tot}_{\rm CO}(\rm lnm+tail)$ as a function of the upper rotational quantum number $J_{\rm up}$. We separately address the case $T_{\rm CMB}(z=0)$ and $T_{\rm CMB}(z=6)$, and we assume two different values for the Mach number (\mach$=10,\,5$). 
In all cases, and for all the rotational transitions, $R_{\rm CO, J}<1$. More precisely, for \mach$=5$ the pure lognormal density distribution can account only for $\approx 20\%-10\%$ ($J_{\rm up}=1-9$) of the CO emission model including the tail. If \mach$=10$, $R_{\rm CO, J}\approx 60\%-10\%$ ($J_{\rm up}=1-9$, $z=0$) and $R_{\rm CO, J}\approx 60\%-40\%$ ($J_{\rm up}=1-9$, $z=6$).
Both at $z=0$ and $z=6$, we note a clear decreasing trend of $R_{\rm CO, J}$ with $J_{\rm up}$, highlighting the strong contribution of the dense tail gas to the high-$J$ lines that have increasingly high excitation temperatures and critical densities \citep{carilli2013}. 
Hence, for a given density distribution, the emission from high-$J$ CO lines is boosted for warmer kinetic temperatures.

This is exactly what we obtain at fixed Mach numbers, where the drop of $R_{\rm CO, J}$ at high-$J$ transitions is steeper at $z=0$ than at $z=6$. This is because the mean kinetic temperature of the molecular gas (i.e. for $N_H>10^{21.5}\, \rm{cm^{-2}}$) is lower at $z=0$ ($\langle T_K \rangle =39\, \rm{K}$), than at $z=6$ ($\langle T_K \rangle =46\, \rm{K}$). 
The increase of the Mach number from \mach$=5$ to \mach$=10$ boosts $R_{\rm CO, J}$, i.e. the emission of the lognormal-distributed gas. The reason is that for large Mach numbers the lognormal distribution becomes wide (eq. \ref{lognormaleq}) and hence a non-negligible fraction of the gas is compressed in high density, albeit not in gravitationally bound, structures. Such turbulent density enhancements are transient, and might not survive long enough time to allow formation of CO and $\rm H_2$ molecules (see later in Sec. \ref{cosled_section}).

\subsection{The CO SLED}
\label{cosled_analysis}
\begin{figure*}
\centering
\includegraphics[scale=0.4]{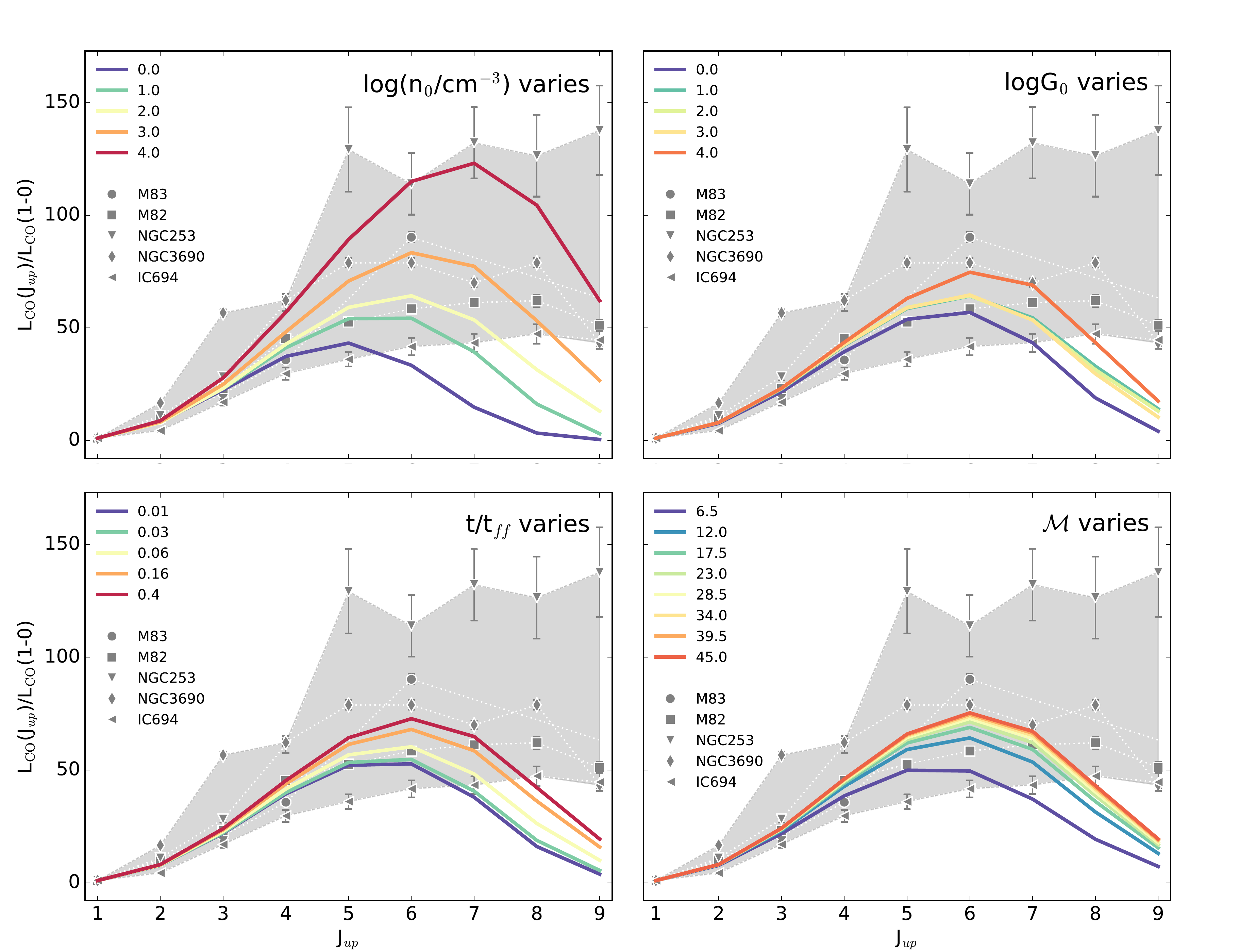}
\caption{The coloured solid lines represent the CO SLEDs, normalised to the CO(1--0) transition, obtained from our modelling at $z=0$, when varying $\rm n_0$, \g, \tff, and \mach~(top-left, top-right, bottom-left, bottom-right panels, respectively). The shaded region highlights the variation of the observed CO SLEDs for starburst galaxies in the nearby Universe \citep{mashian2015}. \label{COSLED_parameters}}
\end{figure*}
Observations of the CO SLEDs can be used as a tool to constrain the properties of molecular gas and to link them to the star formation process, both from single GMCs \citep[e.g.][]{pon2016, lee2016, indriolo2017} and on galactic scales \citep[e.g.][]{rosenberg2015, mashian2015, lu2017, pozzi2017}. Any model that aims at predicting and interpreting the CO emission must be able to reproduce the relative strength of the various lines under different ISM conditions. In Fig. \ref{COSLED_parameters} we investigate how the resulting CO SLED from the fiducial cloud (Tab. \ref{fiducialgmc}) is influenced by variations of (i) the mean density, $n_0$, (top left) (ii) FUV Habing flux, $G_0$, (top right) and the density PDF shape parametrised by (iii) $t/t_{ff}$ (bottom left) and \mach~(bottom right). Given that we will compare our results with local observations of starburst galaxies in the local Universe \citep{mashian2015}, in this Section we adopt \cloudy~runs with present-day CMB temperature. For the reasons explained later, we finally analyse the effects of metallicity variations separately. 

\subsubsection{Mean density effect}
As already mentioned in Sec. \ref{introduction} different CO rotational transitions trace gas with different properties. While low-$J$ rotational transitions ($J_{up}\leq3$) arise from diffuse ($n=10^2-10^4\, \rm{ cm^{-3}}$), cold ($T_K=10-20\,\rm{K}$) molecular ISM, higher-$J$ transitions are excited in denser ($n=10^5-10^6\, \rm{ cm^{-3}}$) and warmer ($T_K=50-600\,\rm{K}$) gas \citep{kaufman1999}. 
In Fig. \ref{COSLED_parameters} we plot the resulting CO SLEDs, normalised to the CO(1--0) transition, for different mean density $n_0$; the values of the other parameters are kept fixed to the fiducial ones. As expected, the peak of the CO SLED rise and shifts from $J_{up}=5\rightarrow7$ with increasing $n_0$.
 
This does not come as a surprise, as the population of high-$J$ CO levels -- set by the competition of collisional excitation and radiative de-excitation -- increases with increasing mean gas density. This eventually boosts the emission of high-$J$ CO lines, and shifts the peak of the CO SLED towards larger $J$ \citep[see e.g.][]{weiss2007, dacunha2013, narayanan2014}.

\subsubsection{FUV radiation effect}
For high-$J$ CO lines, large gas temperatures are required to populate the corresponding rotational levels. An increase of the FUV fluxes at the GMC surface produces warmer gas and can ultimately boost the peak of the normalised CO SLED. Recall that CO(1--0) traces gas that is colder than that in which high-$J$ CO lines are excited.
Note however that, in spite of a large (four dex) variation in $G_0$, the CO SLED peak increases only by $\approx 1.5$ times (upper right panel  of Fig. \ref{COSLED_parameters}. As emphasised also by \citet{kaufman1999}, the temperature in the CII/CI/CO transition layer -- directly influencing the CO emission -- has a weak dependence on the Habing flux. A larger $G_0$ also forces such boundary to move deeper in the cloud.

\subsubsection{Density PDF shape effects}
In the lower left panel of Fig. \ref{COSLED_parameters} we plot the CO SLED dependence on the cloud evolutionary time \tff. As discussed in Fig. \ref{ratio_tail}, the emergence of the high-density tail in the more evolved stages boosts the emission of all CO lines, and more noticeably of the high-$J$ ones. This explain the shift-and-increase effect of the CO SLED peak at larger \tff.
A similar effect is also produced by large Mach numbers (lower right panel). As already explained a larger \mach~ causes a increase in the standard deviation of the of lognormal density distribution, thus allowing the gas to achieve larger densities with a non-negligible probability. This ultimately enhances the emission of high-$J$ CO lines. 

For comparison, in Fig. \ref{COSLED_parameters} we plot in grey the range covered by the observed CO SLEDs (up to $J_{up}=9$) of five nearby starburst galaxies extracted from the sample of \citet{mashian2015}. We do not expect that any of our single cloud model can, alone, reproduce the observed CO SLED on global galactic scales. In fact, the galaxy-integrated CO SLEDs results from the overlap of many different GMCs which have different illumination, density, and temperature conditions. 

Nevertheless the comparison of our simulated CO SLEDs with those observed allows us to highlight three points:
\begin{itemize}
\item Even though it is out of our scope to reproduce in detail the CO SLED shapes of the galaxies in the sample, the predictions for $J_{up}\leq 6$ reproduce well the observed trends.
\item The model under-predicts the luminosity of $J_{up}\geq 7$ lines. This does not come as a surprise, as we have not included shocks in our treatment. Shocks in GMCs are known \citep[][]{pon2012,pon2015,pon2016} to dissipate their energy primarily through CO rotational transitions. In particular, $J_{up} \geq 7-8$ lines come often from shocked gas, and are typically brighter than those predicted by PDR models.
\item Our model assumes a fixed CRIR and this might affect the shape of the CO SLED. For example, the CO SLED of NGC 253, in which the CRs flux is $100-1000$ times the MW one \citep{bradford2003}, is not reproduced by any combination of parameters shown in Fig \ref{COSLED_parameters}. In the Appendix \ref{appendix_CRs} we address in detail the effect of the CRIR variation on the CO line luminosity.
\end{itemize}

\subsubsection{Metallicity effect}
The analysis of the CO SLED presented in Fig. \ref{COSLED_parameters} has been performed at solar metallicity, as we were comparing our results with observations of local starburst galaxies whose $Z\approx Z_{\odot}$.
This is likely not the case of high-$z$ sources \citep[$Z<Z_{\odot}$][]{pallottini2014}, which are the primary target of this work. In what follows we will discuss the impact of $Z$ variations on the CO SLEDs and we will use the observations of multiple CO emission lines in the N159W region of the Large Magellanic Cloud (LMC) \citep[][]{lee2016} as benchmark.

N159W is one of the three prominent GMCs \citep[$M_{\rm H2, N159W}\approx 10^5\rm{M_{\odot}}$][]{fukui2015} in the N159W complex in the LMC \citep[$d_{\rm LMC}\approx50$ kpc, ][]{schaefer2008} whose physical conditions have been extensively studied at multiple wavelengths \citep[][and references therein]{lee2016}. Recently, \citet{lee2016} presented a coherent analysis of the CO and fine structure ([CII] 158$\mu m$, [OI] 145$\mu m$, [CI] 370$\mu m$) line emission to assess the properties of the molecular gas in N159W. N159W is a perfect target to test our model at low metallicities for two reasons: (i) the size\footnote{The area of the map in Fig. 8 of \citet{lee2016} is $\approx 150\times 150\, \rm arcsec^2$. Note, however, that this must be considered as an upper limit on the area of the region from which the CO emission is observed. In fact, only those pixel for which the $S/N>5$ were considered by \citet{lee2016} when computing the total luminosity quoted in their Tab. 3.} of the observed region ($\approx 150\, \rm arcsec$ corresponding to $\approx 35 \,\rm{pc}$) is comparable to the diameter of our fiducial GMC ($30\,\rm{pc}$), and (ii) \citet{lee2016} performed a PDR and shock analysis of N159W CO SLED up to the CO(12--13) transition with which we will compare our results.

\begin{figure}
\includegraphics[scale=0.42]{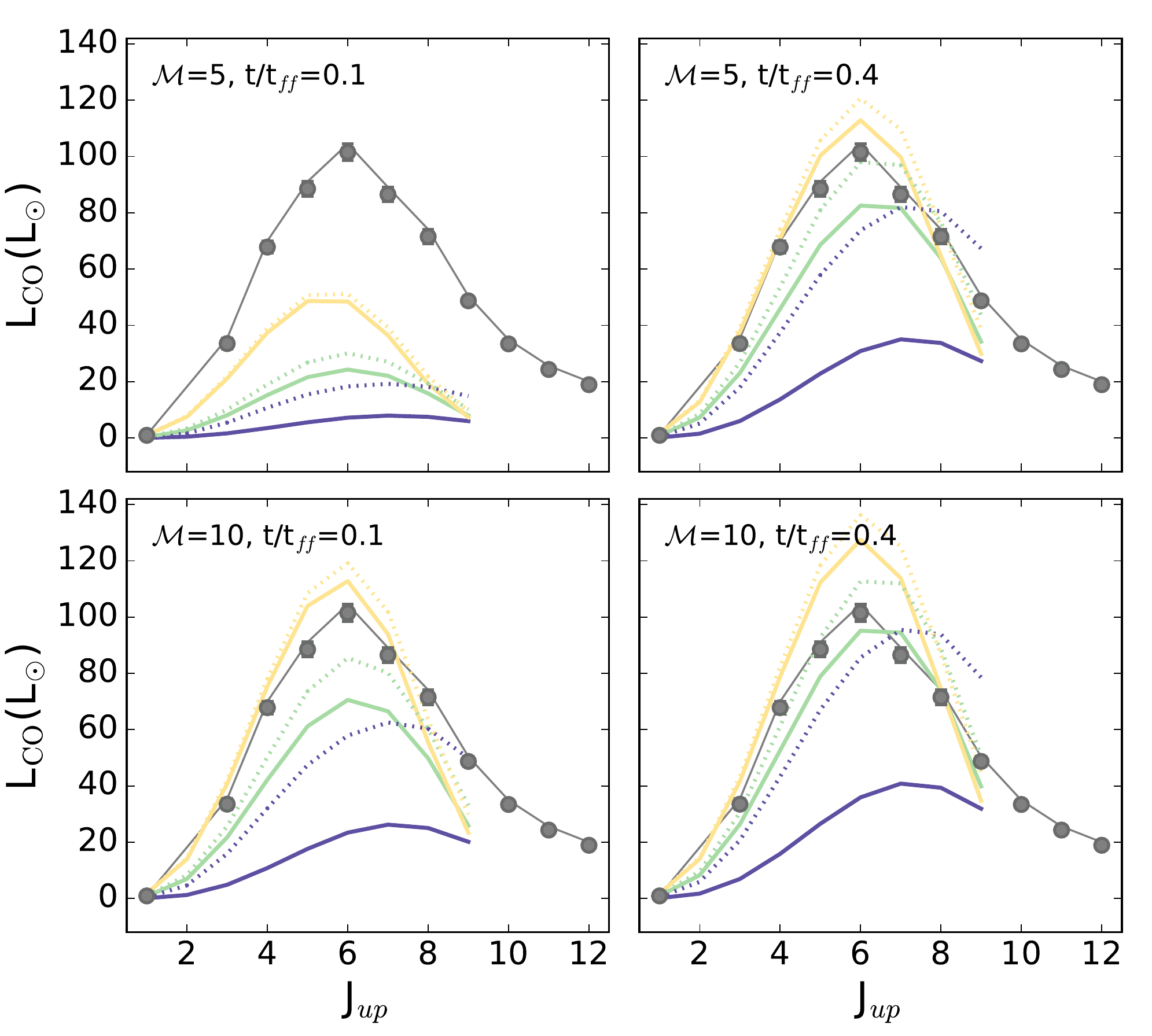}
\caption{The combined effect of the variation of $Z=0.1,\,0.5,1 \, Z_{\odot}$ (purple, green, yellow lines, respectively) and ${\rm log\,}G_0=2,\,1.5$ (dotted, solid lines) on the CO SLED. We consider 4 different models characterised by \mach$=5,\,10$ (upper and lower row, respectively), and \tff$=0.1,\,0.4$ (left and right columns, respectively). The grey points represent the observed CO line emission from the N159W GMC in the LMC \citep{lee2016}. \label{lmc}}
\end{figure}

In Fig. \ref{lmc} we plot the CO SLED as a function of $Z$ assuming two different FUV fluxes at the GMC surface ($\rm{log}\,G_0=1.5,\,2$). We note that the luminosity of all CO lines decreases with $Z$. The effect of the $G_0$ variation (solid and dotted lines in Fig. \ref{lmc}) is negligible at $Z=Z_{\odot}$, but it increases at lower metallicities. This is because, as pointed out by e.g. \citet{chevance2016}, a low metal (and dust) abundance results in less shielding. The FUV photons penetrate deeper into the cloud producing thicker PDRs and smaller CO cores. 
In Fig. \ref{lmc} we show that there is a positive correlation between $L_{\rm CO}$, \mach~ and/or \tff . This is in line with what discussed in the previous sections for the $Z=Z_{\odot}$ runs.
In their PDR analysis of metal FIR lines, \citet{lee2016} found that the best-fit is obtained for ${\rm log}G_0\approx1.8-2.1$. Moreover, from the study of the CO-traced molecular gas they conclude that $T_{K}=153-754\,\rm K$ and $n_{\rm H2}\approx 10^3$. The PDR models used in their analysis, however, fail to explain the CO observations. Their conclusion is that the CO-emitting gas is excited by something other than UV photons, possibly shocks. Our model, including the internal density structure of the GMC is instead successful in matching the data. In fact, we find that models with $Z=0.5\,\rm{Z_{\odot}}$ and ${\rm log}G_0=1.5$ can reproduce the CO SLED up to $J_{up}=8$ if \mach$=10$ and \tff$=0.4$. 

\section{Galaxy simulations}\label{co_highz}
We apply the above CO-emission model to \textit{post-process} a recently produced zoom-in simulation described in \citet[][]{pallottini2017b}. Below we briefly summarise its main features.  

Starting from cosmological initial conditions, we have used a modified version of the Adaptive Mesh Refinement code \textlcsc{RAMSES} \citep{teyssier:2002} to carry out a zoom-in simulation of a $z\sim 6$ dark matter (DM) halo of mass $\sim10^{11} \msun$. In the zoomed-in region the gas has mass resolution of $10^4 \msun$, and dynamics is followed down to spatial scales of $\simeq 30\,{\rm pc}$.
Stars are formed from molecular hydrogen, whose abundance is computed on the fly using the non-equilibrium chemistry code \textlcsc{KROME} \citep{grassi:2014mnras} which is conveniently coupled to our customised version of \textlcsc{RAMSES}.  The thermal and turbulent energy content of the gas is modelled following to e.g. \citet{agertz:2015apj}. As detailed in \citet{pallottini2017}, stellar feedback includes supernovae, winds from massive stars and radiation pressure. Stellar energy inputs and chemical yields depend both on time and stellar populations; the feedback prescription accounts for energy losses inside the GMC (albeit the density structure modelling introduced here has not yet been included).

The selected DM halo hosts ``Alth{\ae}a'', a galaxy characterised by a stellar mass $M_\star\sim 10^{10} \msun$, a mean gas surface density $\langle \Sigma_{gas} \rangle =220 \, \rm M_{\odot}\, yr^{-1}$, and a $\rm SFR\sim 100\, \msun {\rm yr}^{-1}$ at $z\sim6$. Alth{\ae}a features a SFR-stellar mass relation compatible with observations \citep[e.g.][]{jiang2016}, is in agreement with the Schmidt-Kennicutt relation \citep{krumholz2012}, and has a [\CII] emission $\log (L_{\rm CII}/{\rm L}_{\odot}) \simeq 8.3$, slightly lower than the one expected from the local [\CII]-SFR relation \citep{delooze2014} and compatible with some high-z galaxy upper-limits \citep{schaerer2015}.

\subsection{Luminosities of individual simulated clouds}\label{coupling_subgrid}
\begin{figure}
\centering
\includegraphics[scale=0.35]{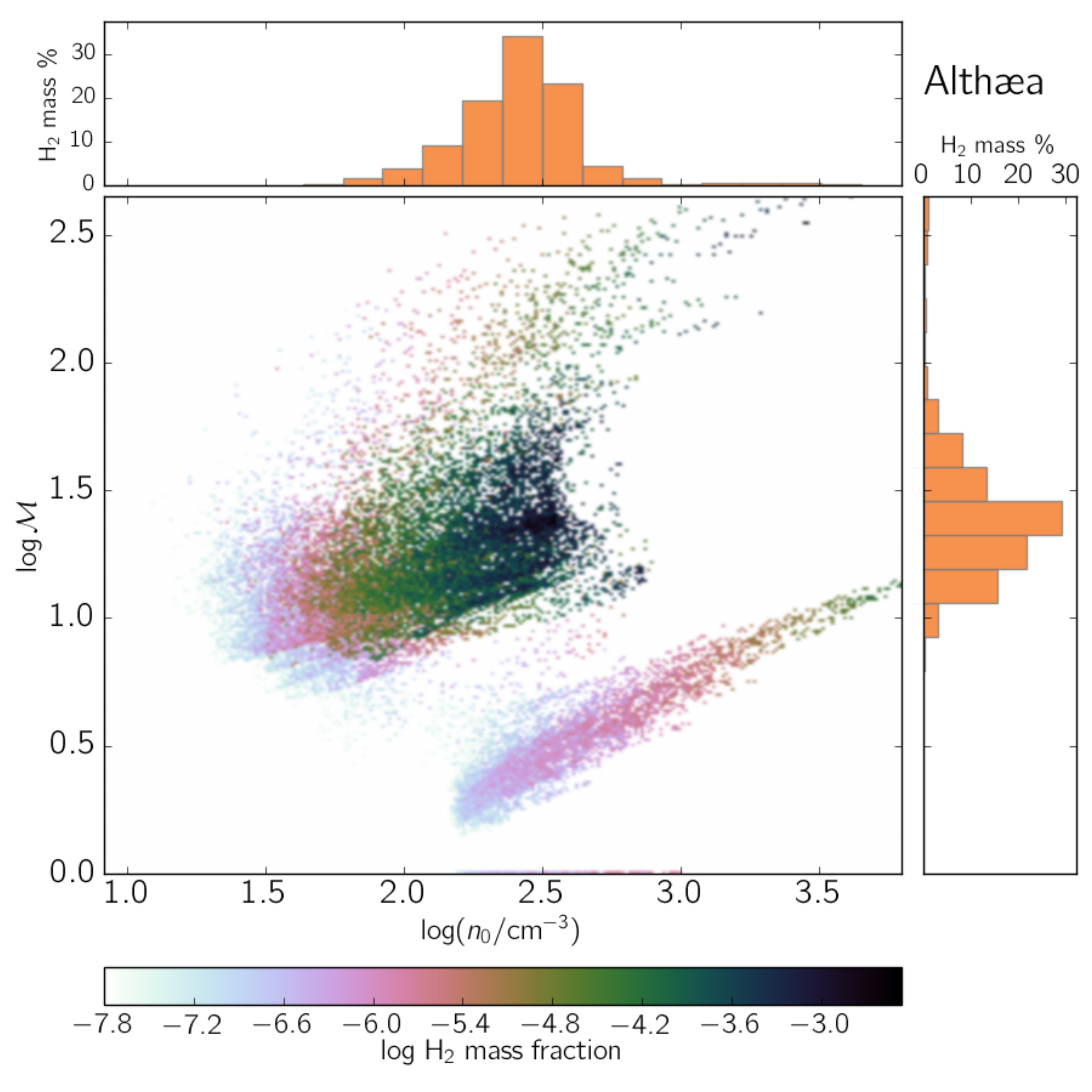}
\caption{Mach number vs density in Alth\ae a at $z=6$. The relation is plotted using the $\rm H_2$ mass weighted PDF. The projections on the $n$ (\mach) axis are shown as horizontal (vertical) insets. \label{mach_althaea}}
\end{figure}
As outlined in Fig. \ref{model_outline}, and discussed in Sec. \ref{outline}, our CO model needs \tff, \mach, $n_0$, $R_{\rm GMC}$, $G_0$, and $Z$ as inputs.
In Fig. \ref{mach_althaea} we plot the $\rm H_2$ mass weighted PDF of \mach -$n_0$ relation in Alth{\ae}a, considering a $z=6$ snapshot of the simulation. 
In Alth{\ae}a the density PDF peaks at ${n_0}\approx 300\,\rm cm^{-3}$.  In addition there is a small fraction of very dense gas ($n_0>1000 \,\rm cm^{-3}$) -- the low \mach\, diagonal stripe. This part of the PDF describes the virtually metal-free gas in which $\rm H_2$ production proceeds via gas-phase reactions rather than on dust grain surfaces and can survive only if self-shielded by a high density.
The Mach number has a relatively wide distribution (see the inset of the Figure) with a pronounced peak at \mach$\approx30$. This high level of turbulence is mostly supported by momentum injection associated with radiation pressure onto dust around massive stellar clusters and supernova explosions. It is a consequence of the very high star formation rate per unit area which in Alth{\ae}a is about 1000 times higher than in the Milky Way.    
\begin{figure}
\centering
\includegraphics[scale=0.65]{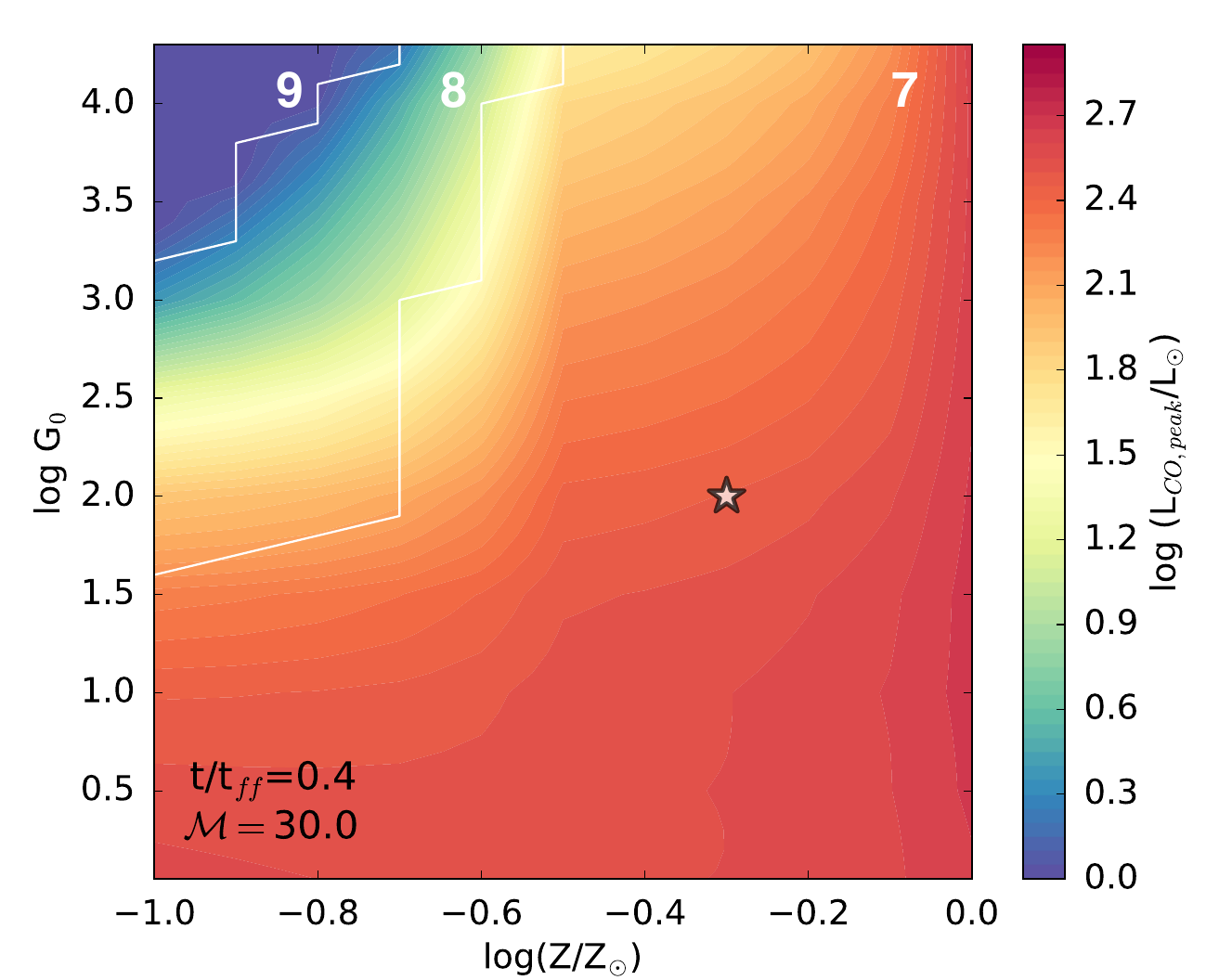}
\caption{Luminosity of the CO transition at peak of the CO SLED (J$_{up}$ indicated on the contours) as a function the $G_0$ and $Z$. The contours are obtained for \tff=$0.4$, assuming \mach$ =30$  and $n_0=300\, \rm{cm^{-3}}$. \label{peakluminosity}}
\end{figure}
To guide the following interpretation, in Fig. \ref{peakluminosity} we plot as a function of \Z~and \g~ the luminosity of the brightest transition (indicated by the white contours) from a single GMC characterised by \mach$=\langle \mathcal{M}\rangle_{\rm Althaea} =30$ and $n_0=\langle n_{0} \rangle_{\rm Althaea}=300$. Considering that $\langle Z \rangle_{\rm Althaea} = 0.5 \, Z_{\odot}$ and $\langle G_0 \rangle_{\rm Althaea}= 100$, we expect a typical $L_{CO}(7-6)\approx 10^2\,\rm L_{\odot}$ from such a cloud, which has a mass of $1.4\times10^5\, \rm{M_{\odot}}$.
The CO SLED peaks at CO(7--6) transition for all $Z$ if ${\rm log G_0}<1.5$, otherwise, if ${\rm log\,G_0}>1.5$ the peak shifts towards higher $J_{\rm up}$ with decreasing metallicity. 
\begin{figure*}
\centering
\includegraphics[width=0.8\textwidth]{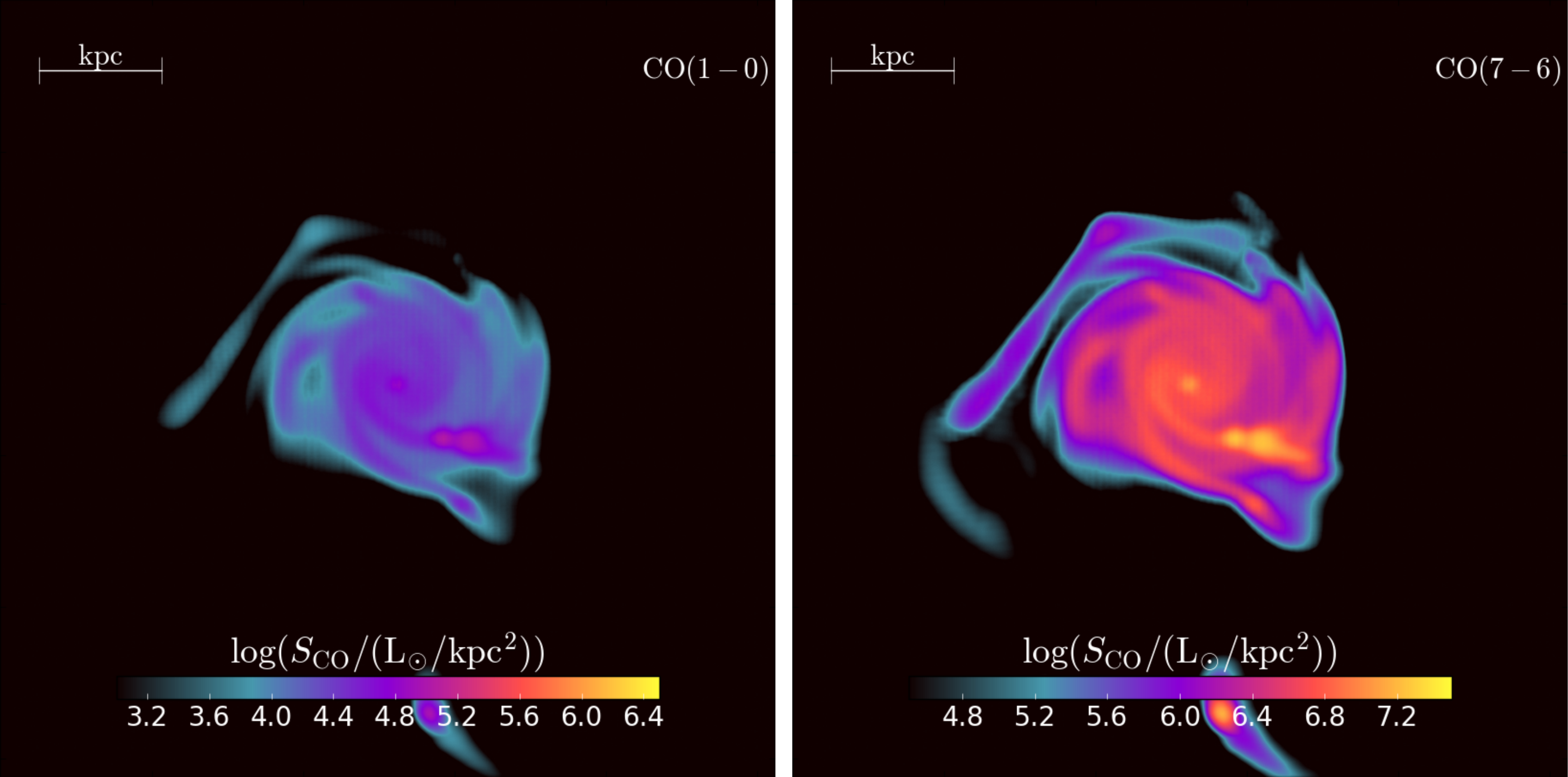}
\caption{CO(1--0) and CO(7--6) surface brightness maps of Alth{\ae}a. Note the different scale of the colourbars in the two panels. \label{maps}}
\end{figure*}
\subsection{CO emission from Alth{\ae}a} 
Figure \ref{maps} shows the morphology of the CO(1--0) and CO(7--6) emission in Alth{\ae}a. We select these two lines among all the CO rotational transitions because: (i) the CO(1--0) enters in the calculation of the CO-to-$\rm H_2$ conversion factor, $\alpha_{\rm CO}$, which we will discuss in detail in Sec. \ref{alphaco_section}) and, (ii) the CO(7--6) is the most luminous transition of the Alth{\ae}a CO SLED (see Sec. \ref{cosled_section}) and it is observable with ALMA from $z\approx 6$.

As extensively discussed in \citet{pallottini2017b}, Alth{\ae}a features a clearly defined, even though rather perturbed, spiral disk of radius $\approx 0.5$ kpc, embedded in a lower density ($n\simeq0.1\, {\rm cm^{-3}}$) medium. The CO emission traces the disk where indeed most of the $\rm H_2$ mass resides. Both the CO(1--0) and the CO(7--6) maps show an enhanced emission clump along the spiral arms. 
The peak of the CO(1--0) and CO(7--6) surface brightnesses are coincident (${\rm log} (S_{\rm CO(1-0)}/{\rm L_{\odot} \,kpc^{-2}})\approx 5.0$, and ${\rm log}(S_{\rm CO(7-6)}/{\rm L_{\odot} \, kpc^{-2}})\approx 7.2$, respectively). Not surprisingly, they are co-located with a high-density ($n\approx 10^{3}\, \rm cm^{-3}$) clump where also the $\rm H_2$ 17.03 $\mu m$ and [\CII] lines reach their maximum surface brightness \citep{pallottini2017b}.

\begin{figure}
\centering
\includegraphics[width=0.5\textwidth]{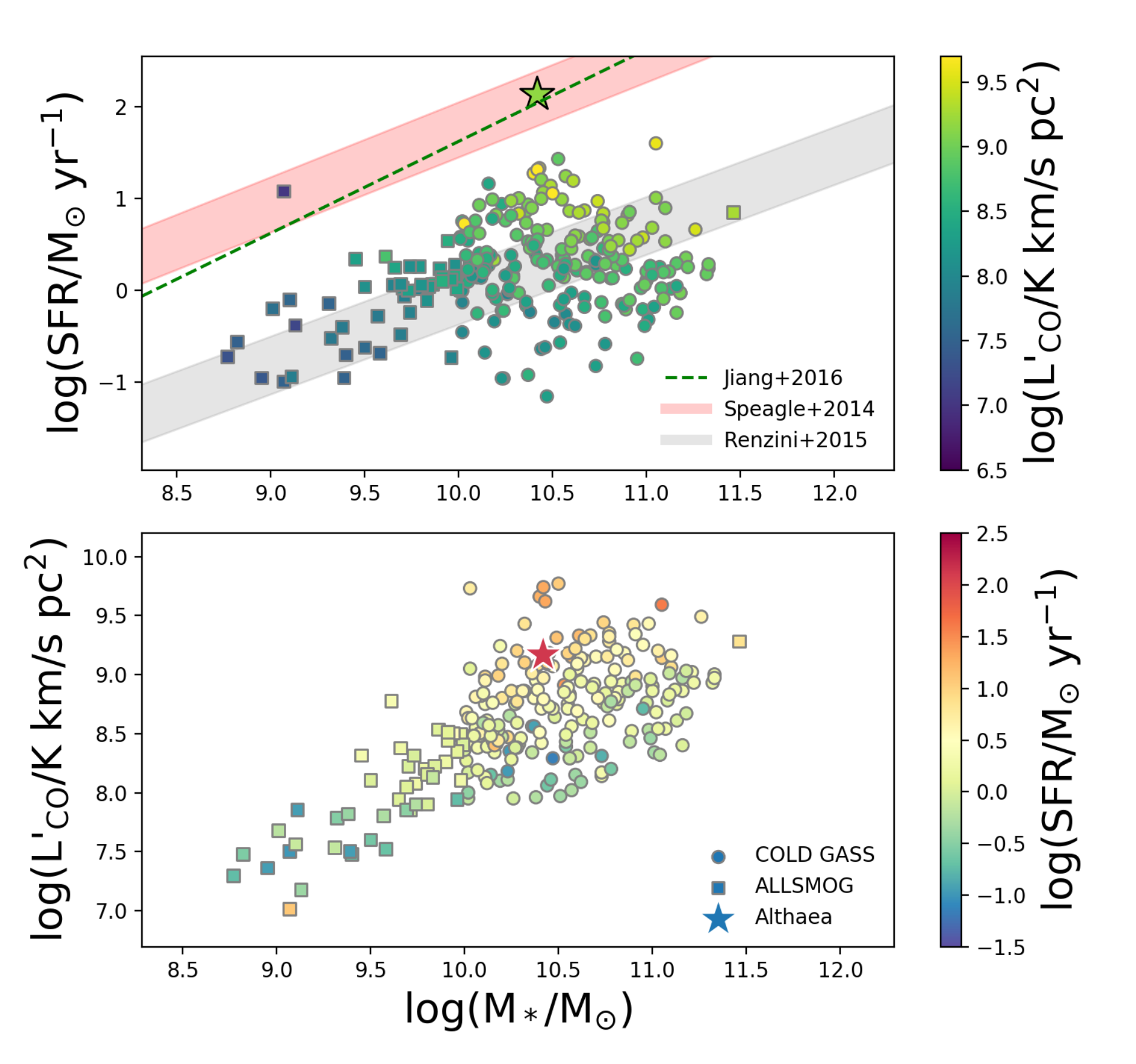}
\caption{Upper panel: The COLD GASS \citep{saintonge2011} (circles) and ALLSMOG \citep{cicone2017} (squares) samples are plotted in the $M_{*}$-SFR plane and colour-coded as a function of the CO(1--0) luminosity. The gray (red) shaded region indicates the location of the star-forming main sequence (MS) at $z\approx0$ \citep{renzini2015}, \citep[$z=6$,][]{speagle2014}; the green dashed line is the best fit obtained from observations in LBGs/LAEs at $z\approx 6$ \citep{jiang2016}. Alth{\ae}a's position is marked with a star. Lower panel: $L'_{\rm CO}-M_{*}$ relation for COLD GASS, ALLSMOG and Alth{\ae}a, color coded as a function of SFR.  \label{MS_lco}}
\end{figure}

To understand how the total CO(1--0) luminosity of Alth{\ae}a compares with local observations, in the upper panel of Fig. \ref{MS_lco} we plot the $M_{*}-$SFR relation for galaxies in the COLD GASS \citep{saintonge2011}, and ALLSMOG \citep{cicone2017} samples, as well as the location of Alth{\ae}a in the same plane. The points are colour-coded in $L'_{\rm CO}$. We also highlight the evolution of the star-forming ``main-sequence'' (MS) from $z=0$ \citep{renzini2015} to $z=6$ \citep{speagle2014, jiang2016}.

Alth{\ae}a nicely falls on the MS at $z=6$ \citep{speagle2014}. Its $M_{*}-$SFR relation is in agreement with that recently found by \citet{jiang2016} in $z\approx6$ LAEs and LBGs characterised by stellar ages $>100\, \rm{Myr}$. As shown in the lower panel of Fig. \ref{MS_lco}, the CO(1--0) luminosity of Alth{\ae}a is $L'_{\rm CO}=10^{9.17}\, \rm{K \, km\, s^{-1} \, pc^{2}}$ ($L_{\rm CO}=10^{4.8}\rm \, L_{\odot}$), and it is comparable to that of galaxies with the same stellar mass at $z\approx0$. 

The specific star formation rate (${\rm sSFR=SFR/}M_{*}$) of Alth{\ae}a is higher than that of MS-galaxies at $z=0$, i.e. its SFR is larger than that of galaxies with comparable $M_{*}$ (see the color code of the Alth{\ae}a symbol in the lower panel of Fig. \ref{MS_lco}). 
Therefore the CO(1--0) luminosity per unit SFR in Alth{\ae}a is lower than that of $z\approx 0$ MS galaxies.
\subsection{The CO SLED in Alth{\ae}a}
\label{cosled_section}
\begin{figure}
\centering
\includegraphics[width=0.42\textwidth]{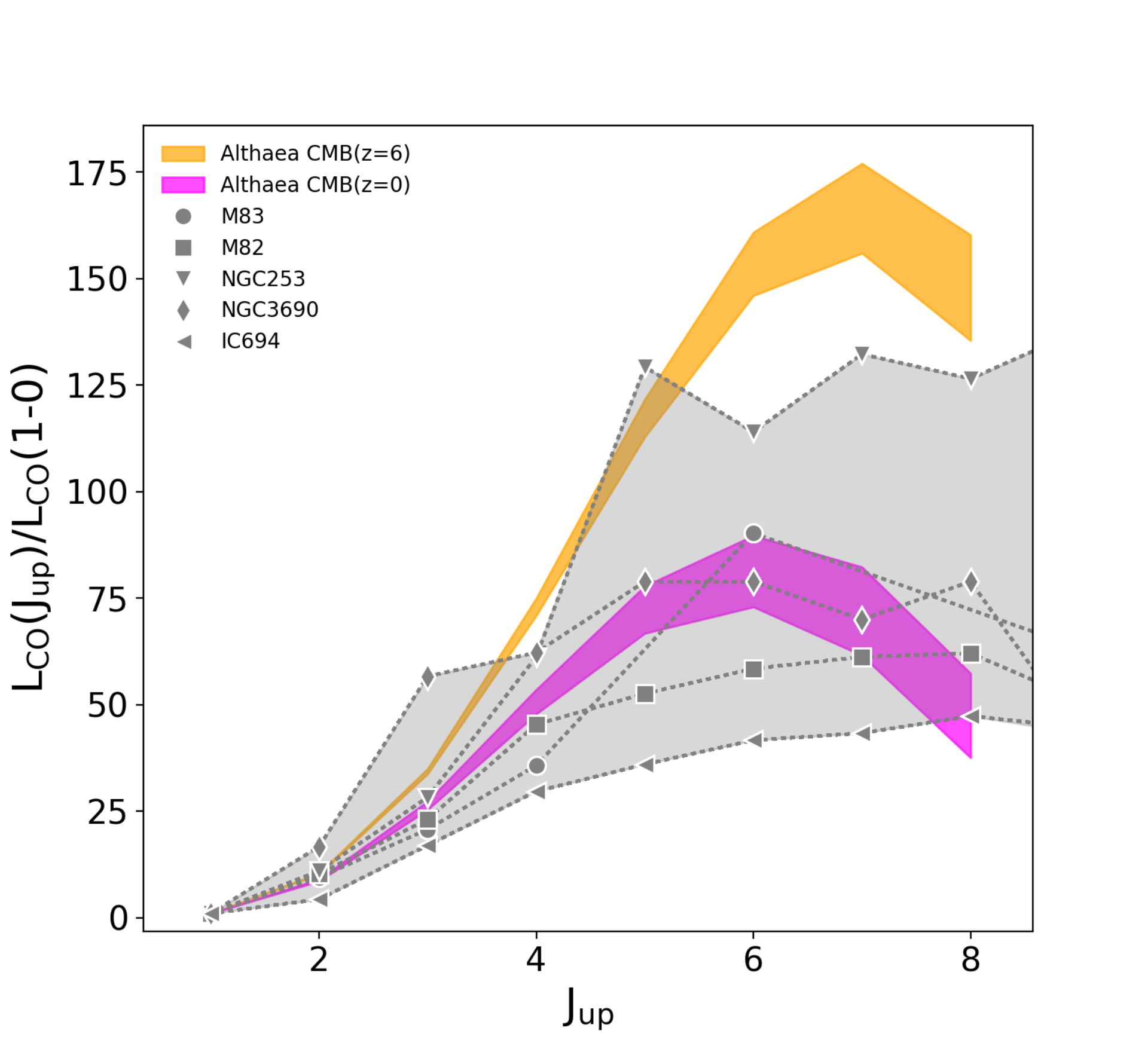}
\caption{The CO SLED in Alth{\ae}a normalised to the CO(1--0) transition. In orange we plot the actual CO SLED, while the magenta shaded region represents the CO SLED of Alth{\ae}a considering the CO emission obtained with RT calculations with T$_{\rm CMB}(z=0)$. The grey points represents the CO SLEDs of the five starburst galaxies in the \citet{mashian2015} sample. The shaded area in the simulated SLEDs corresponds to variation of \tff~ in the range $[0.01, 0.4]$.. \label{cosled_althaea_mashian}}
\end{figure}

The CO SLED is a unique tool to infer the properties of molecular gas, even if the simultaneous effect of the various parameters (e.g. the gas density, the FUV field, the gas metallicity) often makes its interpretation challenging. This is the reason why, in Sec. \ref{cosled_analysis}, we separately discussed the effect of each of the relevant parameters entering in our modelling. Here, we will refer to the results of that analysis to interpret the CO SLED of Alth{\ae}a plotted in Fig. \ref{cosled_althaea_mashian}.

The CO SLED of Alth{\ae}a includes the effects of CMB background radiation at $z=6$. To isolate the effects of the CMB, we have also produced a case in which the CMB temperature is fixed at its present-day ($z=0$) value. The latter peaks at $J=6$ and is consistent with the range spanned by the observed CO SLEDs in a sample of local starburst galaxies \citep{mashian2015}.
The higher average GMC temperature due to the warmer CMB at $z=6$, shifts the SLED peak at $J=7$; in addition, the galaxy is about 2 times more luminous in the brightest line available if located at $z=6$ compared to the $z=0$ case.

According to \citet{dacunha2013} the observed flux of the $J$-th line ($S^{\rm obs}$) against the CMB over the intrinsic one can be expressed as:
\begin{equation}\label{dacunha_cmb}
\frac{S^{\rm obs}_J}{S^{\rm intrinsic}_J} = 1- \frac{B_{\nu}[T_{\rm CMB}(z)]}{B_\nu[T_{ex}]},
\end{equation}
where $\nu$ is the rest frame frequency of the transition, and $B_{\nu}$ is the black body spectrum at  temperature $T$.
Using this equation, the ratio of the observed versus intrinsic CO(7--6) flux from $z=6$ (substituting $T_{\rm CMB}(z=6)=19.1$ K and $T_{ex}=T_{k, \rm GMC, \, z=6}=46\, \rm K$) is $\approx0.8$. 

Then, we can estimate the ratio of the observed CO(7--6) from $z=6$ over that at $z=0$. This can be done by computing the intrinsic CO(7-6) luminosity in LTE \citep[][see eq. 4,5]{obreschkow2009}, using $T_{ex}=39\, \rm K$ ($T_{ex}=46\, \rm K$) for $z=0$ ($z=6$). This yields $S^{\rm obs}_J(z=6)/S^{\rm obs}_J(z=0)=0.8\, S^{\rm int}_J(z=6)/S^{\rm int}_J(z=0) \approx 1.3$.

Qualitatively the CO(7--6) suppression due to the CMB (eq. \ref{dacunha_cmb}) is compensated by the increased temperature of GMCs at high-$z$.
However the ratio as estimated in LTE ($\approx 1.3$) is slightly lower than what we find ($\approx 2.3$) from our model (Fig. \ref{cosled_althaea_mashian}).  The origin of such discrepancy is due to a combination of the following factors: (i) \cloudy~calculations show that the excitation temperature inside clouds is not uniform but has a varying spatial profile; (ii) the predicted CO luminosity in Alth{\ae}a arises from a collection of emitting clouds with different densities; (iii) the presence of an external stellar radiation field, \g, induces deviations from the LTE regime implicitly assumed by eq. \ref{dacunha_cmb}.

In Fig. \ref{cosled_althaea_mashian} the Alth{\ae}a SLEDs are represented with shaded areas which highlight the variation of the CO line luminosity as a function of \tff. We let \tff~vary in the range $[0.01, 0.4]$, that causes an increase of $L_{\rm CO}$ by a factor $\approx 1.5$. The impact of the tail is relatively small because the average Mach number of molecular cells in the simulation is $\approx 30$ (Fig. \ref{mach_althaea}), which implies  $\sigma\approx 2.0$ (see eq. \ref{lognormaleq}). In this case the median density of the mass-weighted PDF is $\rho_{50 \%}=n_0\times e^{-0.14}$ \citep[see e.g. Tab. 1 in][]{girichidis2014}. Given that the typical number density of molecular cells is $n_0\approx 300 \rm{cm^{-3}}$, this means that 50\% of the mass actually resides in structures with densities above $\rho_{50 \%}=260 \rm \rm{cm^{-3}}$ even without including the the contribution of the tail. Given that for Alth{\ae}a the \tff~parameter has a minor impact on the CO luminosity, in what follows we fix \tff$=0.1$.

However, we re-emphasise that, as discussed in Sec. \ref{mvir_lco_section}, this statement holds true only in chemical equilibrium, i.e. if the timescale for $\rm H_2$ (and CO) formation \citep[$t_{\rm H2}^{-1}\propto \mathcal{R}n$, with $\mathcal{R}\approx 10^{-17} \rm \, cm^3\,s^{-1}$][]{jura1975} is shorter than the typical lifetime of turbulent density enhancements.
A simple estimate can be obtained as follows. Let us consider a clump with $n=10^3\rm{cm^{-3}}$, corresponding to a typical scale $L_e\approx 1\rm \, pc$ (see eq. \ref{clump_radius}). At such density, the H$_2$ formation timescale is $t_{\rm H2}\approx 10^{14}\, \rm s$ \protect\citep[see][for a detailed calculation]{liszt2007, glover2010}.  Let us assume that the clump lifetime is the eddy turnover time ($t_e=L_e/v_e$). Using Larson's law, we know that the turbulent velocity scales as $v_e\propto L_e^{1/3}$ hence, $v_e=\mathcal{M}c_s(L_e/(2R_{\rm GMC}))^{1/3}$. Substituting $\mathcal{M}\approx 30$, and $c_s=0.3\, \rm{km\, s^{-1}}$ we obtain $t_e\approx10^{13}\, \rm{s} < t_{\rm H2}$.
This simple estimate shows that it can be difficult for $\rm H_2$ (and CO) to form in a purely turbulent environment and gravitational collapse might be needed to keep the overdense regions bound.
% the presence of a well-developed density tail due to gravitationally collapsing gas. 
Finite lifetime of clumps and non-equilibrium chemistry is typically not considered in CO emission calculations that explore a wide range of physical conditions \citep[e.g.][]{kazandjian2016}, while it is accounted for in single cloud simulations \citep[e.g.][]{glover2010, shetty2011}.
This is an interesting aspect that is worth to be investigated in future work.

\subsection{CO-to-$\rm H_2$ conversion factor}
\label{alphaco_section}
The so-called CO-to-$\rm H_2$ conversion factor is defined as the ratio of the molecular gas mass to the CO(1--0) line luminosity:
\begin{equation}
\label{alpha_co_eq}
\alpha_{\rm CO} \equiv \left(\frac{M_{\rm H_2}}{L'_{\rm CO}}\right)  \rm M_{\odot}(K\, km\, s^{-1}\, pc^2)^{-1}.
\end{equation}

Observationally, $\alpha_{\rm CO}$ is determined by combining independent measurements of $\rm H_2$ gas mass with the detection of the CO(1--0) line. There are three methods to infer the molecular mass and/or column density: (1) assume that GMCs are in virial equilibrium, and derive the $\rm H_2$ mass from the CO line width \citep[e.g.][]{larson1981, solomon1987}, (2) assume a constant dust-to-gas ratio and use the dust continuum emission, possibly combined to HI measurements, to infer $N_{\rm H_2}$, e.g. \citet[][in resolved Galactic GMCs]{pineda2008} and \citet[][for extragalactic studies]{leroy2011, magdis2011, sandstrom2013}; (3) through $\gamma$-ray emission induced by cosmic ray interactions with $\rm H_2$ molecules \citep[e.g.][]{padovani2009}.
In the MW disk the conversion factor is fairly constant, $\alpha_{\rm CO} = 4.3\pm 1.3 \rm \, M_{\odot} \,(K\, km\, s^{-1}\, pc^2)^{-1}$  \citep{bolatto2013}  on large ($\approx$kpc) scales.

In recent years, a number of observational studies have provided evidence for at least two physical regimes where $\alpha_{\rm CO}$ departs from the MW value. The first deviation is observed in sources with high-surface density and/or warm molecular gas such as mergers, and starburst galaxies where the conversion factor is lower than MW one \citep[e.g.][]{yao2003, tacconi2008, papadopoulos2012}. To first order high gas temperatures, and gas surface densities ($\Sigma_{\rm H2}$), yield brighter CO emission at fixed molecular mass, thus decreasing the conversion factor \citep{narayanan2011, narayanan2012}.

A deviation in the opposite direction is instead observed in low-metallicity galaxies. In these sources $\alpha_{\rm CO}$ is larger than in the MW \citep[e.g.][]{bolatto2008, leroy2011}. 
This is due to the low C and O abundances and, most importantly, to the low dust-to-gas ratios which prevent an efficient shielding against FUV dissociating photons\footnote{We recall that, contrary to H$_2$, CO cannot self-shield.}. The reduced CO abundance then produces a fainter luminosity. The actual value of $\alpha_{\rm CO}$ in high-redshift ($z\geq6$) galaxies is far from being firmly constrained. On the one hand, high-$z$ galaxies are more compact, dense, and star-bursting than the MW; all these facts lead to lower $\alpha_{\rm CO}$ values. On the other hand, if their metallicity is sub-solar, one would expect higher $\alpha_{\rm CO}$ ratios. 

\begin{figure}
\centering
\includegraphics[width=0.4\textwidth]{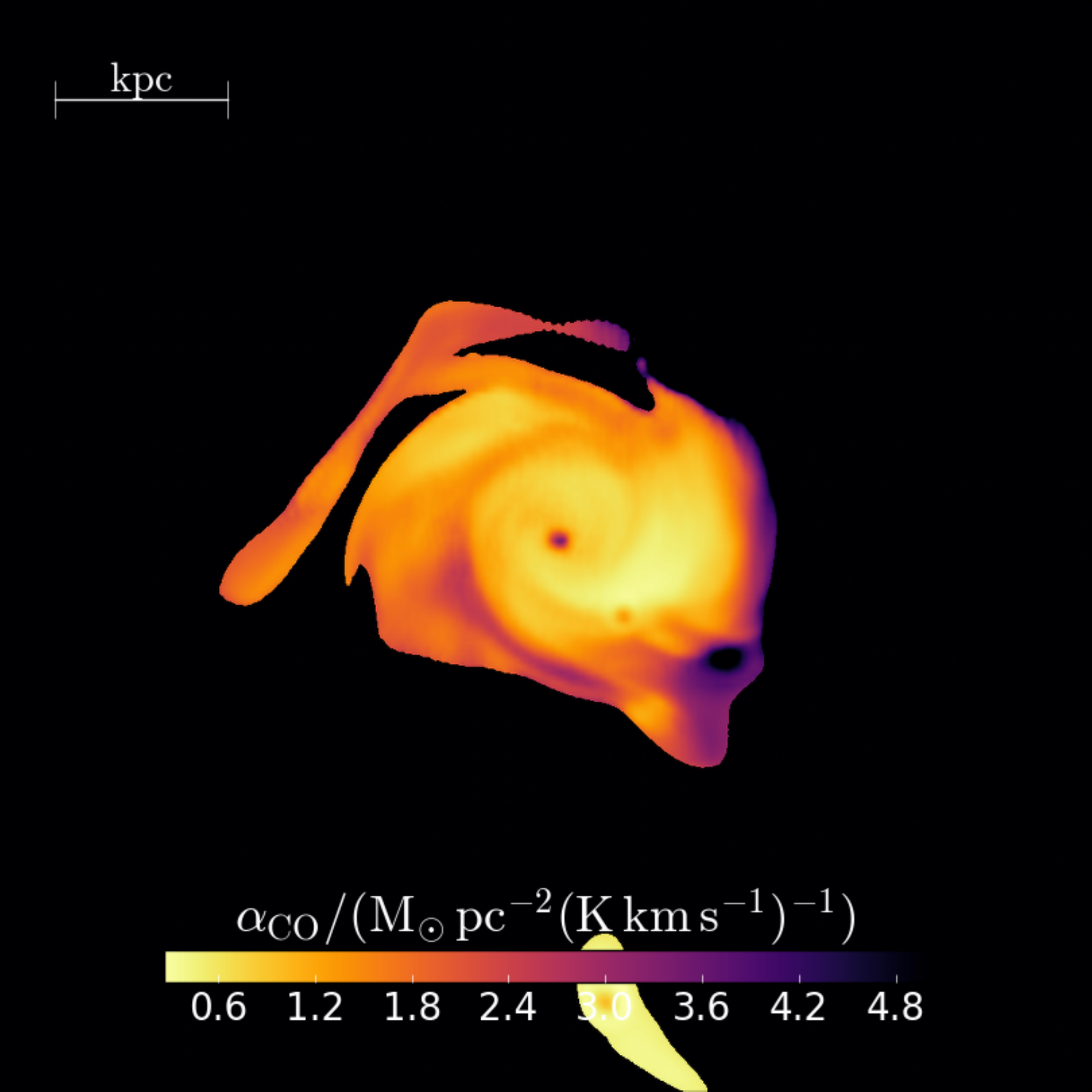}
\caption{The map of $\alpha_{\rm CO}$ in Alth{\ae}a. \label{alphacoalthaea}}
\end{figure}

In Fig. \ref{alphacoalthaea} we plot the $\alpha_{\rm CO}$ map of Alth{\ae}a. The map is obtained by dividing the CO(1--0) surface brightness by the corresponding $\rm H_2$ mass surface density.  The mass of molecular hydrogen in each cell is:
\begin{equation}
M_{\rm H2, cell}= f_{\rm H_2, GMC} \, M_{\rm gas, cell},
\end{equation}
where $f_{\rm H2, GMC}$ is the molecular fraction obtained with the sub-grid model (see Appendix \ref{appendix_molecular_fraction} for details). 
The mean value (standard deviation) of the CO-to-$\rm H_2$ in Alth{\ae}a is $\langle \alpha_{\rm CO} \rangle = 1.54 \pm 0.9 \rm \, M_{\odot}/(K\, km\, s^{-1}\, pc^{2})$. Even though \mach~and $n_0$ have similar relative variations in the disk of Alth{\ae}a, i.e. $ \sqrt{\langle\mathcal{M}^2\rangle}/\langle\mathcal{M}\rangle \approx  \sqrt{\langle n_0^2\rangle}/\langle n_0 \rangle \approx 0.2$, most of the dispersion of $\alpha_{\rm CO}$ is due to the density fluctuation. 
The $\alpha_{\rm CO}$ is rather constant trough the disk and this is in agreement with what found by \citet{sandstrom2013} that pointed out that the radial profile of $\alpha_{\rm CO}$ in spiral galaxies is mostly flat. However, \citet{sandstrom2013} found also a slight decrease of the conversion factor towards the center of the observed galaxies. In Alth{\ae}a we do not see such a trend.

As pointed out previously, $\alpha_{\rm CO}$ can be influenced by the gas surface density $\Sigma_{\rm H_2}$, by the strength of the Habing field, by the gas metallicity \Z, and also by the CRIR.
  
Recently, \citet{clark2015, glover2016} performed detailed numerical simulations of turbulent molecular clouds that cover a wide range of metallicities, strength of the interstellar radiation field (ISRF), and cosmic ray ionisation rate, to investigate the impact of these parameters on the conversion factor. They find that $\alpha_{\rm CO}$ increases with decreasing metallicities, in agreement with observations in the nearby Universe \citep[e.g.][]{bolatto2008, leroy2011}. Increasing the ISRF and the CRIR produces different effects on the well-shielded clumps, where CO survives effectively, and in the diffuse interclump medium, where CO is instead dissociated effectively. Hence, even for high values of the ISRF, the integrated intensity from dense clumps increases owing to the heating of the gas. This is especially relevant in the case of model clouds characterised by high-density and high turbulent velocity dispersion, for which the conversion factor drops close to the MW value, even when the local ISRF is 100 times the fiducial (i.e. $G_0=1.7$) value. 

The warm temperature of the molecular gas, sustained by the CMB at $z=6$, and the high level of turbulence and gas surface densities of Alth{\ae}a, are likely the two main reasons pushing the Alth{\ae}a CO-to-$\rm H_2$ conversion factor below that of the MW, despite the sub-solar metallicity, $\langle Z \rangle=0.5\, \rm{Z_{\odot}}$, and ${\rm SFR}\approx100\,\rm{M_{\odot}\, yr^{-1}}$ featured by Alth{\ae}a. 
As a caveat, we also note that in our model we totally neglect any type of stellar feedback on GMCs \citep[e.g.][]{gorti2002, vallini2017, decataldo2017}, which may affect the density field of GMCs, and ultimately increase the conversion factor.

Finally, we compare our inferred conversion factor with the one resulting from the best fit formula, $X_{\rm CO}/({\rm cm^{2}\, K\, km \,s^{-1}})= 6.3 \times 10^{19} \alpha_{\rm CO}/({\rm M_{\odot} K\, km\,s^{-1}\, pc^2})$, obtained by 
\citet{narayanan2012}. They simulate the hydrodynamic evolution of both isolated and merging disk galaxies finding that $X_{\rm CO}= (1.3 \times 10^{21})(Z/Z_{\odot})^{-1} (\Sigma_{\rm H_2}/(\rm{M_{\odot}\,pc^{-2}}))^{-0.5}$. If we substitute  $\Sigma_{\rm H_2} =f_{\rm H_2, GMC} \langle \Sigma_{\rm gas} \rangle \approx 200 \, \rm{M_{\odot}\,pc^{-2}}$ and $Z=0.5 \, Z_{\odot}$ we obtain $\alpha_{\rm CO}=2.9 \, \rm M_{\odot}/(K\, km\, s^{-1}\, pc^{2})$ in agreement within a factor $\approx 2$ with our results.

\subsection{ALMA observability}
In Sec. \ref{cosled_section} we have shown that the peak of the Alth{\ae}a CO SLED coincides with the CO(7--6) line. As CO transitions with $J_{\rm up}\geq6$ fall in the ALMA bands from $z>6$, in what follows, we use the ALMA Sensitivity Calculator to compute the observing time required to detect the CO(7--6) line from (a galaxy similar to) Alth{\ae}a. The total CO(7--6) luminosity of Alth{\ae}a is $L_{\rm CO(7-6)}=10^{7.1} \, \rm L_{\odot}$, i.e. $\approx 1/16$ of its [\CII] luminosity. If we assume the line width to be equal to that of [\CII] observations in LBGs at $z\approx6-7$ \citep[$\rm FWHM\approx 150 \, \rm km \, s^{-1} $][]{maiolino2015, pentericci2016, knudsen2016, bradac2017}, this yields a peak flux density $F^{p}_{\rm CO(7-6)}\approx0.179\, \rm{mJy}$\footnote{The peak flux is related to the luminosity in solar units by the following equation \citep{obreschkow2009}
\begin{equation}
\frac{F^p_{\rm CO}}{\rm Jy} = 961.5\,\frac{L_{\rm CO}}{L_{\odot}} \left(\frac{D_L}{\rm Mpc}\right)^{-2}\left( \frac{\nu_{o}}{\rm GHz}\right)^{-1} \left(\frac{FWHM}{\rm km\, s^{-1}}\right)^{-1},\nonumber
\end{equation}
where $D_L$ is the luminosity distance to the source in Mpc, and $\nu_{o}$ is the observed frequency in GHz.}. The ALMA full-array observing time required to detect (resolve over $1/3\times \rm FWHM$) the CO(7--6) line with a signal-to-noise ratio S/N=5 is $\approx 13$ h ($\approx 38$ h), i.e. it would be challenging but doable within the maximum observing time ($\leq50\, \rm h $) allowed for ALMA regular programs.
This is in line with what found by \citet{munoz2013}, which pointed out that ALMA observations of high-$J$ CO can be performed with reasonable integration times only in those $z\approx6$ galaxies which, like Alth{\ae}a, are chemically evolved ($Z_{\rm Althaea}>0.1\, Z_{\odot}$) and UV-bright ($M_{\rm UV, Althaea}\approx -22$).

\section{Conclusions}
\label{conclusions}
In this paper we have studied the CO emission properties of galaxies at the end of the Reionization Epoch. 
First, we have developed a semi-analytical model that, given the internal density field of a GMC and the impinging FUV flux (\g) at the cloud surface, computes the CO line emission. The density PDF of the GMC is set both by turbulence (parametrised by the Mach number \mach) and self-gravity (parametrised by \tff). The radiative transfer is performed with \cloudy~and includes $z=6$ CMB radiation.
The model takes the mean gas density $n_0$, \mach, \tff, \g~ as inputs, and it returns the luminosity of the first 9 CO rotational transitions.
We validated the model with local observations, demonstrating its capability to reproduce:
\begin{itemize}
\item[-] the $M_{vir}$-$L'_{\rm CO}$ relation 
\item[-] the observed CO excitation in local SB galaxies
\item[-] the CO SLED of the low-metallicity ($Z\approx0.5\,\rm{Z_{\odot}}$) molecular cloud N159W in the LMC. 
\end{itemize}

We used our validated CO emission model to post-process a cosmological zoom-in simulation of a prototypical  $z\approx6$ galaxy ($M_\star\sim 10^{10} \msun$, $\rm SFR\sim 100\, \msun {\rm yr}^{-1}$),  Alth{\ae}a. We have made the following additional assumptions: (i) we adopt a constant CRIR (cfr. Appendix \ref{appendix_CRs} for a discussion on the effects of the CRIR variation); (ii) the dust-to-gas ratio scales linearly with metallicity; (ii) all clumps within a GMCs are exposed to the same (external) FUV flux predicted by the simulation; (iv) photoevaporation of clumps affecting the density field and the lifetime of GMCs has been neglected. 

The key results can be summarised as follows:
\begin{itemize}

\item[1.] The CO emission traces the innermost disk of Alth{\ae}a where most of the $\rm H_2$ mass resides. 

\item[2.] The CO(1--0) luminosity of Alth{\ae}a is comparable to that of main sequence galaxies with similar stellar mass at $z\approx0$. As the MS evolves with redshift, and the sSFR increases, this means that the CO(1--0) per unit SFR in Alth{\ae}a is lower than measured in local galaxies. 

\item[3.] The CO-to-$\rm H_2$ conversion factor is $\langle \alpha_{\rm CO} \rangle = 1.54 \pm 0.9 \rm \, M_{\odot}/(K\, km\, s^{-1}\, pc^{2})$, with little spatial variation throughout the disk. The dispersion is primarily introduced by density variations in the Alth{\ae}a disk. 

\item[4.] The maximum of the CO(1--0) and CO(7--6) surface brightnesses are colocated in the disk ($S_{\rm CO(1-0)}\approx 10^5 \,{\rm L_{\odot} \,kpc^{-2}}$, and $S_{\rm CO(7-6)}\approx10^{7.2} \,{\rm L_{\odot} \, kpc^{-2}}$, respectively).

\item[5.] The suppression of the observed CO luminosity due to the CMB at high-$z$ is compensated by the increased temperature of GMCs. The net result is both an increase of the CO SLED excitation, and a shift of the peak at higher $J$.

\item[6.] The peak of the Alth{\ae}a CO SLED coincides with the CO(7--6) transition, and $L_{\rm CO(7-6)}=10^{7.1}\, \rm L_{\odot}$ i.e. $\approx 1/16$ of the [\CII] luminosity. This is due to the relatively high surface density of Alth{\ae}a and to the warm temperature ($T_k\approx 45 \rm\, K$) of the GMCs. To resolve the CO(7--6) line with a S/N=5 an ALMA observing time of $\approx 38\, \rm h$ is required. 
\end{itemize}

\section*{Acknowledgments}
We thank G. Ucci, D. Cormier, R. Maiolino, M. Bothwell for useful discussions. We thank the anonymous referee for the valuable feedback that increased the clarity of the paper. AF acknowledges support from the ERC Advanced Grant INTERSTELLAR H2020/740120. ES acknowledges support from the Israeli Science Foundation under Grant No. 719/14.

%%%% APPENDIX %%%%%%
\bibliographystyle{mnras}
\bibliography{COhighz}
\appendix

\section{Time evolution of the density PDF of a GMC}
\label{sec:ghirichidismodel}
Following \citet{girichidis2014}, we first consider a homogeneous sphere, with initial density $\rho_0$, collapsing under its own gravity. The density at any later time $t$ can be well approximated by
\begin{equation}
\label{eq:rhoff}
\rho=\rho_0\left[1-\left(\frac{t}{t_{\rm ff}}\right)^2\right]^{-2}\;,
\end{equation}
where
\begin{equation}\label{app_tff}
t_{\rm ff}\equiv\sqrt{\frac{3\pi}{32G\rho_0}}
\end{equation}
is the free-fall time of the sphere.

Due to the conservation of the probability density, one can calculate the evolved density PDF as
\begin{equation}
P_V(\rho,t)=P_V(\rho_0,0)\frac{\text{d}\rho_0}{\text{d}\rho}\;,
\end{equation}
where $\rho_0$ needs to be expressed as a function of $\rho$ and $t$ inverting eq. (\ref{eq:rhoff}).

\section{Effect of varying the cosmic ray ionisation rate}
\label{appendix_CRs}
In our simulations we assume $\zeta_{\rm CR} = 2\times10^{-16} \rm \, s^{-1}$ \citep{indriolo2007}. This value is $\approx 20$x higher than the CRIR ($\zeta_{\rm CR,0} = 10^{-17} \rm \, s^{-1}$) generally adopted as default in many PDR calculations \citep[e.g.]{glover2012, bisbas2015}. The actual CRIR value in Alth{\ae}a $-$ and in general in high-$z$ galaxies $-$ is highly uncertain, owing to e.g. the unknown magnetic field in the ISM of high-$z$ galaxies.

A simple estimate of the CRIR can be obtained assuming a linear scaling with the SFR, as the main source of CRs is  the Fermi acceleration in supernova (SNe) remnants, and the rate of SNe is related to the rate at which stars form. In Alth{\ae}a the SFR$\approx100 \rm \, M_{\odot}\, yr^{-1}$, thus, a linear CRIR-SFR scaling would imply $\zeta_{\rm CR,\rm Althaea} \approx \zeta_{\rm CR,0} \times {\rm SFR} \approx 10^{-15}\rm \,  s^{-1}$ i.e. at most a factor of 5 greater than the value adopted in our \cloudy~simulations. 
 \begin{figure}
 \centering
\includegraphics[scale=0.4]{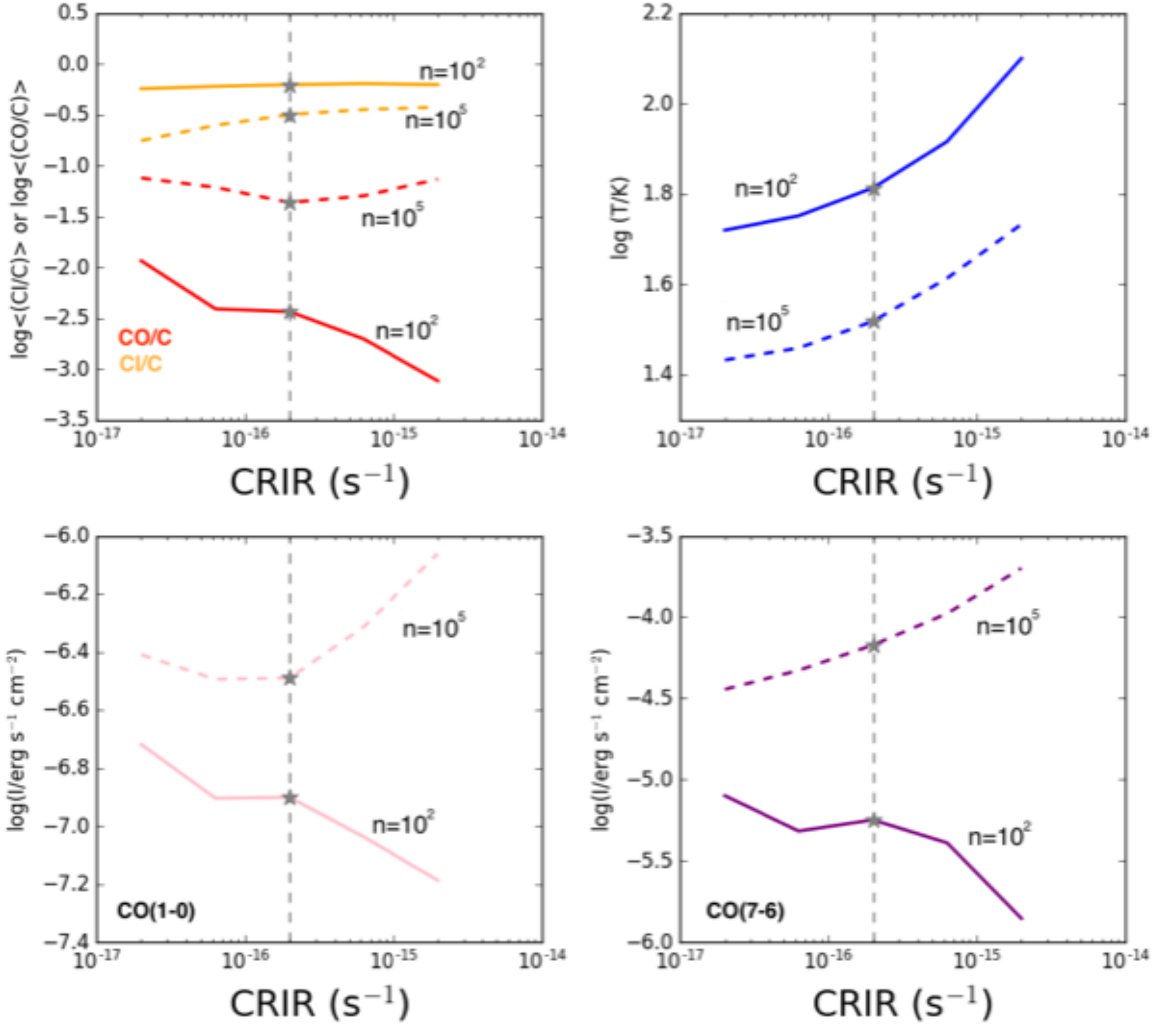}
\caption{\textit{Upper left panel:} variation of the average $\langle n_{CO}/n_{C} \rangle$ and $\langle n_{CI}/n_{C} \rangle$ ratios for $n=10^2\rm \, cm^{-3}$ (solid lines) and $n=10^5\rm \, cm^{-3}$ (dashed lines), as a function of the CRIR. \textit{Upper right:} variation of the gas temperature at $N_H=10^{22} \rm \,cm^{-2}$. \textit{Bottom left:} variation of the CO(1--0) emissivity. \textit{Bottom right:} variation of the CO(7--6) emissivity. The dashed vertical line highlights the CRIR adopted in this work. \label{test_CRIR}}
\end{figure}
To test the impact of the CRIR variation on the CO (and CI) abundance, gas temperature, and CO line emissivity, we run a set of \cloudy~models with CRIR in the range $[2\times 10^{-17}-2\times 10^{-15}] \rm \, s^{-1}$, keeping $\rm G_0=100$ and $Z=0.5\rm \,  Z_{\odot}$ fixed to the fiducial values of Althaea.
In Fig. \ref{test_CRIR}, we show the variation of the $\langle n_{CO}/n_{C} \rangle$ and $\langle n_{CI}/n_{C} \rangle$ ratios (upper left panel), gas temperature (upper right) at $N_H=10^{22} \rm \,cm^{-2}$, CO(1--0) emissivity (bottom left), CO(7--6) emissivity (bottom right), that we obtain in two different set of CLOUDY runs at fixed gas number density $n=10^2\rm \, cm^{-3}$ and $n=10^5\rm \, cm^{-3}$. 
The $\langle n_{CO}/n_{C} \rangle$ and $\langle n_{CI}/n_{C} \rangle$ ratios are obtained by averaging the $\langle n_{CO}(r)/n_{C} \rangle$ and $\langle n_{CI}(r)/n_{C} \rangle$ returned by \cloudy~as a function of the depth ($r$) into the gas slab. These values are proportional to the total fraction of carbon locked up in CO and CI, respectively.

We note that at low densities ($n=10^2\rm \, cm^{-3}$, solid lines) $\langle n_{CO}/n_{C} \rangle$ decreases with increasing CRIR, while the opposite is true for $\langle n_{CI}/n_{C} \rangle$. This trend is expected as an increase of CRIR produces more He$^+$. This ultimately boosts the dissociative charge transfer reactions of the CO molecules with He$^+$ ions. In shielded regions the ionised carbon produced by these reactions is then converted into neutral carbon, boosting the CI/C ratio. Note, however, that a 100x variation in the CRIR causes a drop of $\approx1/10$ in $\langle n_{CO}/n_{C} \rangle$. 
As expected, the gas temperature, sampled at $N_H=10^{22} \rm \,cm^{-2}$, increases with CRIR, as the heating provided by CRs is proportional to the cosmic rate ionisation rate. 
The CO(1-0) and the CO(7-6) emissivities decrease of $\approx 0.6\rm \,dex$, as the boost in the gas temperature is not enough to compensate for the CO abundance decrease.

At high densities ($n=10^5\rm \, cm^{-3}$, dashed lines) the situation is slightly different. The $\langle n_{CO}/n_{C} \rangle$ ratio decreases only of a factor $\approx 2$ between CRIR $[2\times 10^{-17}-2\times 10^{-16}] \rm \, s^{-1} $ and $\langle n_{CO}/n_{C} \rangle$ remains almost constant, with a very shallow increase above $\zeta_{\rm CR}=1\times 10^{-15} \rm \, s^{-1}$. 
%This is in line with the findings of \citet{bisbas2015}, namely an increase in the CO abundance for CRIR $>5 \times 10^{-15} \rm \, s^{-1}$ at $n=10^4 \rm \, cm^{-3}$. 
The higher temperature makes the CO(1--0) and CO(7--6) emissivity increase of $\approx 0.4-0.5\, \rm dex$, respectively, for CRIR varying between $[2\times 10^{-17}-2\times 10^{-15}]\, \rm s^{-1}$, and compensate the small CO drop. This in line with \citet{bisbas2015} (see their Fig. 3), and \citet{glover2016} which note that very dense clumps remain CO-bright despite the increase of the cosmic rate ionisation rate. 

As pointed out by \citet{glover2016}, accounting for the influence of turbulence on the density field of GMCs when assessing the impact of the CRIR variation on the CO emission is pivotal, as a significant GMC mass fraction may be located in regions with a mean density higher than the volume-weighted mean density of the cloud. 
\citet{glover2016} show that, in this case, a variation of CRIR by a factor of 100 has only a minor effect on e.g. the values of the $\alpha_{\rm CO}$ conversion factor. Given that, and the results shown in Fig. \ref{test_CRIR}, we expect that uncertainties on the CRIR of a factor of 5 cause a negligible variation in the CO line luminosity. This is especially true for the CO(7-6) line (owing to the high critical density of the transition, $n_{\rm cr}=4.5\times10^5 \rm \, cm^{-3}$) for which we provide our ALMA predictions.

\section{Molecular fraction from the sub-grid model}
 \label{appendix_molecular_fraction}
 \begin{figure}
\centering
\includegraphics[scale=0.4]{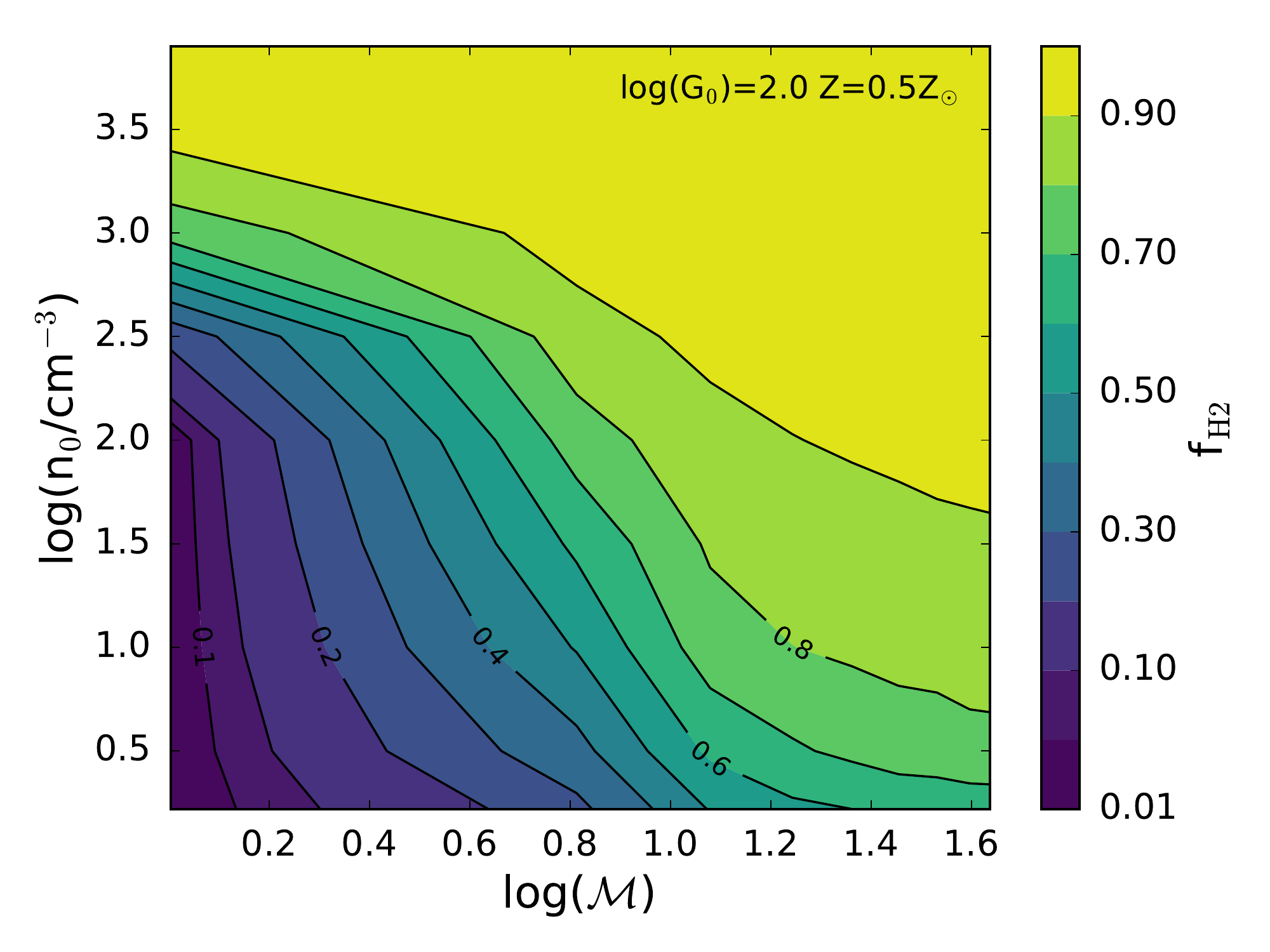}
\caption{Molecular fraction as a function of $n_0$ and \mach. We fix \g~ and \Z~ to those of Alth{\ae}a. \label{H2fraction}}
\end{figure}
 As we are interested in how efficiently the CO luminosity traces the $\rm H_2$ mass, and given that $L'_{\rm CO}$ is computed through the sub-grid model, we need the actual molecular mass returned by the model itself. The molecular fraction from the hydrodynamical simulation ($f_{\rm H2, sim}$) -- that does not account for the internal density structure of the GMCs -- underestimates the molecular mass as $f_{\rm H2, sim} < f_{\rm H2, GMC}$ and hence predicts an artificially low conversion factor. 
 
The molecular fraction from the sub-grid model ($f_{\rm H_2, GMC}$) has been calculated through the following procedure.
Let $f_{\rm H_2}(\rho_i, r)$ be the $\rm H_2/HI$ radial profile returned by \cloudy~for the gas slab of constant density $n_i=\rho_i/m_p$, illuminated by FUV flux $G_0$ at the surface, and characterised by a gas metallicity \Z. We remind the reader that the radius, $r_i$, column density, $N_i$, of the various clumps are set by eq. \ref{clump_radius}.
The $\rm H_2$ mass of each clump of constant density $n_i$ is:
\begin{equation}
m_{\rm H_2}^i = 2 m_p n_i \int_0^{r_i} 4 \pi f_{\rm H_2}(n_i, r) r^2  dr.
\end{equation}
The molecular fraction of each GMC is then:
\begin{equation}
f_{\rm H_2, GMC}=\frac{M_{\rm H2, tot}}{M_{\rm GMC}}= \frac{\sum_i m_{\rm H2, i}}{\sum_i m_{tot}^i}
\end{equation}
In Fig. \ref{H2fraction} we plot $f_{\rm H2, GMC}$ as a function of the mean density $n_0$ and Mach number \mach, assuming \g~ and \Z \,equal to those of Alth{\ae}a. We note that the molecular fraction increases with $n_0$ and \mach. For values $n_0=\langle n_0 \rangle = 10^{2.5} \, \rm cm^{-3}$ and \mach$=30$, i.e. those typical of Alth{\ae}a molecular disk, we obtain $f_{\rm H_2, GMC}\approx 0.9$.

\end{document}